\documentstyle[twocolumn,eqsecnum,aps]{revtex}
\def\dnot#1{#1\llap{/}\relax}
\begin{document}
\draft
\wideabs{
\title{Perturbative nonequilibrium dynamics of phase transitions in an
expanding universe}
\author{Ian D. Lawrie}
\address{Department of Physics and Astronomy, The University of Leeds,
Leeds LS2 9JT, England}
\date{\today}
\maketitle
\begin{abstract}
A complete set of Feynman rules is derived, which permits a perturbative
description of the nonequilibrium dynamics of a symmetry-breaking phase
transition in $\lambda\phi^4$ theory in an expanding universe.  In contrast to
a naive expansion in powers of the coupling constant, this approximation scheme
provides for (a) a description of the nonequilibrium state in terms of its own
finite-width quasiparticle excitations, thus correctly incorporating dissipative
effects in low-order calculations, and (b) the emergence from a symmetric
initial state of a final state exhibiting the properties of spontaneous symmetry
breaking, while maintaining the constraint $\langle\phi\rangle\equiv 0$.
Earlier work on dissipative perturbation theory and spontaneous symmetry
breaking in Minkowski spacetime is reviewed.  The central problem addressed is
the construction of a perturbative approximation scheme which treats the
initial symmetric state in terms of the field $\phi$, while the state that
emerges at later times is treated in terms of a field $\zeta$, linearly related
to $\phi^2$.  The connection between early and late times involves an infinite
sequence of composite propagators.  Explicit one-loop calculations are given of
the gap equations that determine quasiparticle masses and of the equation of
motion for $\langle\phi^2(t)\rangle$ and the renormalization of these equations
is described.  The perturbation series needed to describe the symmetric and
broken-symmetry states are not equivalent, and this leads to ambiguities
intrinsic to any perturbative approach.  These ambiguities are discussed in
detail and a systematic procedure for matching the two approximations is
described.

\end{abstract}

}
\section{Introduction\label{intro}}

It has long been apparent that the hot, dense matter present in the early
universe might undergo a variety of phase transitions \cite{kirzhnits1972}.
In particular, the ``new inflation'' scenario of Linde \cite{linde1982} and
Albrecht and Steinhardt \cite{albrecht1982} (modifying an earlier proposal of
Guth \cite{guth1981}) suggests that a symmetry-breaking phase transition at the
GUT scale would lead to a period of quasi-exponential expansion, perhaps with
highly desirable cosmological consequences (see e.g. \cite{kolb1990}). Variants
of this proposal are still of great current interest.  Amongst cosmologists, the
conventional view is that the state of a quantum field during an inflationary
era is adequately described by classical field theory (except for the purpose
of estimating density perturbations that are taken to originate from small
quantum fluctuations) and, on the basis of classical calculations, that the
new inflation scenario does not work in detail.  However, a fully
quantum-field-theoretic analysis of the dynamics of phase transitions in, say,
grand unified theories, has never been given.

Apart from its intrinsic theoretical interest, a solution of this problem is of
some importance.  If there is a cosmological era in which matter is adequately
described by a spontaneously broken gauge theory, then the consequences of the
associated phase transition need to be correctly understood, whether they lead
to the desired inflationary picture or not. Moreover, the above conclusions
drawn from classical calculations are not necessarily secure. One reason is that
the effective potential which traditionally appears in the classical equations
of motion is not, in principle, an appropriate dynamical tool (for the reasons
discussed in \cite{evans1985} amongst others).  Another is that the value of a
classical field (say, the expectation value of a quantum field) does not
necessarily provide an adequate characterisation of the actual quantum state.
In an influential paper \cite{guth1985}, Guth and Pi argue, for the case of an
unstable free scalar field theory, that at sufficiently late times the quantum
probability distribution evolves in an approximately classical manner, because
the minimal quantum uncertainty is neglible.  However, as these authors
recognise, this does not by any means imply that, say, the expectation value of
the energy density $\langle\rho(\phi)\rangle$ that appears in the field
equations of semiclassical general relativity is well approximated by
$\rho(\langle\phi\rangle)$.  Furthermore, the era of quasiclassical evolution
arises from a growth in the amplitudes of unstable field modes, which may be
considerable (see also \cite{boyanovsky1993}) so that the effects of
interactions may become significant even in a weakly coupled theory.

The purpose of this paper is to derive a set of Feynman rules, by means of which
the nonequilibrium evolution of a quantum field theory may be estimated
perturbatively during the course of a symmetry-breaking phase transition.  One
would like, for example, to be able to solve the semiclassical field equations
\begin{equation}
{\cal G}_{\mu\nu}=\kappa\langle T_{\mu\nu}\rangle\,,
\label{fieldequations}
\end{equation}
where $T_{\mu\nu}$ is the stress tensor of an appropriate quantum field theory,
and the question we address is how the expectation value can be estimated.
Here, we study the example of a self-interacting scalar field, but the methods
we develop can be generalised to the case where this field belongs to the Higgs
sector of a spontaneously broken gauge theory.  For simplicity, we restrict
attention to a conformally coupled field in a spatially flat Robertson-Walker
universe with the line element $ds^2=a^2(t)\left[dt^2-d\bbox{x}^2\right]$, so
that $t$ is conformal time and the spatial coordinates $\bbox{x}$ are comoving.
The action for this theory can be expressed as
\begin{equation}
S(\phi)=\int d^4x\left[\frac{1}{2}\partial_\mu\phi\partial^\mu\phi +
\frac{1}{2}m^2(t)\phi^2 -\frac{\lambda}{4!}\phi^4\right]\,,
\label{action}
\end{equation}
where $m^2(t)=a^2(t)m_0^2$ and $m_0$ is the bare mass of the corresponding
Minkowski-space theory.  With $m_0^2>0$, we would expect the zero-temperature
Minkowski-space theory to exhibit spontaneous symmetry breaking, and we take
the initial state (at a time that we shall call $t=0$) to be a state of thermal
equilibrium at a temperature $1/\beta_0$ which is high enough for the symmetry
to be unbroken.  The expectation value of a Heisenberg-picture operator
${\cal O}(t)$ is then
\begin{equation}
\langle{\cal O}(t)\rangle={\rm Tr}\left[e^{-\beta_0H(0)}{\cal O}(t)\right]
/{\rm Tr}\left[e^{-\beta_0H(0)}\right]\,,
\end{equation}
where $H(t)$ is the Hamiltonian, which depends explicitly on time through the
time-dependent mass $m(t)$.  This choice of an initial state is somewhat
artificial, and is motivated mainly by the fact that it yields a well-defined
problem. It has, in particular, the property that
$\langle\phi(\bbox{x},0)\rangle=0$. This feature will cause considerable
difficulty, since it implies that $\langle\phi(\bbox{x},t)\rangle=0$ at
{\it all} later times also, so it is worth discussing at the outset.  Of course,
this initial state is well-defined, and should evolve in a well-defined manner,
which it is of theoretical interest to investigate.  One may wonder, however,
whether the condition $\langle\phi\rangle=0$ is too special to warrant the
technical effort needed to deal with it, and we want to argue that it is not.
We are supposing that, like the gauge symmetries of more realistic models, the
symmetry $\phi\leftrightarrow -\phi$ is an exact symmetry of nature.  This means
that any field configuration $\phi(\bbox{x})$ is physically indistinguishable
from the configuration $-\phi(\bbox{x})$.  There is therefore no physical
mechanism that can produce different probabilities for these two configurations,
which implies that $\langle \phi(\bbox{x},t)\rangle$ vanishes identically at
every spacetime point $(\bbox{x},t)$. (This does not by any means imply that the
state is spatially uniform, since a quantity such as the energy density, which
is not constrained by symmetry, may perfectly well have a non-uniform
expectation value even when $\langle\phi\rangle=0$.)  In the context of
semiclassical gravity, and in the spirit of the Copenhagen interpretation of
quantum mechanics, it is reasonable to suppose that the density matrix can
depend only on quantities which couple to (and can therefore be ``measured'' by)
the classical spacetime.  It might therefore depend on, say, the energy density
and pressure, which might be spatially non-uniform, but not, independently of
these, on $\phi$, which cannot be measured either by the spacetime or by any
other physical probe.  If $\phi$ belongs to the Higgs sector of a gauge theory,
then these qualitative arguments are superfluous, since a well-known theorem of
Elitzur \cite{elitzur1975} assures us that its expectation value must vanish.

For a variety of reasons, some of which we shall have cause to discuss in detail
later, perturbation theory is a rather limited tool for treating this problem,
but we know of no other approximation scheme which might be used to treat the
single scalar field considered here, or the more realistic models of particle
physics that one might wish to tackle.  The functional Schr\"odinger picture
approach developed in \cite{eboli1988} seems to be restricted in practice to
Gaussian wavefunctionals (and, perhaps, to scalar fields in planar universes).
Motivated by the belief that growing unstable modes make perturbation theory
completely unreliable, Boyanovsky, Holman and de Vega, with several other
collaborators, have studied in considerable detail the case of $O(N)$-symmetric
scalar field theory in the limit $N\to\infty$ (see \cite{boyanovsky1998} and
references therein).  In this limit, the problem can be solved exactly (up to
the numerical solution of a set of integro-differential equations) without the
aid of perturbation theory.  While significant insights can be gained in this
way, the large-$N$ limit is a rather special one (in which, for example,
dissipative effects are absent). Corrections of order $1/N$ and beyond seem to
be intractable, so this approach does not appear to provided the basis for a
non-perturbative treatment of any more realistic models.

In constructing a perturbative means of tackling the problem, we shall need to
draw on earlier work which investigated firstly the possibility of treating
dissipation by describing a nonequilibrium state in terms of its own
quasiparticle excitations and, secondly, how the phenomena normally associated
with spontaneous symmetry breaking can be recovered when $\langle\phi\rangle=0$,
by dealing instead with $\langle\phi^2\rangle$ which is not constrained to
vanish.  The results of this earlier work are summarised in section
\ref{PREVIOUS} below. We find, in particular, that two different perturbative
expansions are needed to describe the symmetrical state which exists at early
times and the broken-symmetry state which exists at later times. Section
\ref{FEYNMANRULES} describes a construction of the path integral which
facilitates this dual description and derives the complete set of Feynman rules.
In section \ref{EOMCON}, we apply these rules to determine, at the lowest
nontrivial order of our approximation scheme, the gap equations for
quasiparticle masses and the evolution equation for $\langle\phi^2\rangle$.
Both this equation of motion and the differential equations obeyed by the
late-time propagators require boundary conditions, which we obtain from the
continuity of appropriate expectation values.  Renormalization of the gap
equation, the equation of motion and the boundary conditions which apply to them
is considered in section \ref{RENORMALIZATION}. Finally, the virtues and
shortcomings of the approximation scheme we propose are discussed in
section \ref{DISCUSSION}.

\section{Summary of previous results\label{previous}}

\subsection{Dissipative perturbation theory\label{dissipation}}

Our calculations of nonequilibrium expectation values are based on the
closed-time-path formalism
\cite{schwinger1961,keldysh1964,mathanthappa1962,chou1985,landsmann1987}. More
specifically, we adopt the path-integral technique described by Semenoff and
Weiss\cite{semenoff1985}, in which Green's functions are obtained from the
generating functional
\begin{equation}
Z(j_a)=\int[d\phi_a]\exp\left[i\bar{S}(\phi_a)+i\int d^4x\,j\cdot\phi\right]\,.
\label{generatingfunctional}
\end{equation}
Here, the single quantum field $\hat{\phi}(\bbox{x},t)$ is represented by three
path integration variables $\phi_a(\bbox{x},t)$ ($a=1, \cdots, 3$), which can be
envisaged as inhabiting three segments of a contour in the complex time plane.
The segment labelled by $a=1$ is associated with time-ordered products.  It runs
from an initial time (which we shall call $t=0$) to $t_f-i\epsilon$, where $t_f$
is the largest time in which we are interested and $\epsilon$ is infinitesimal.
The segment $a=2$, associated with anti-time-ordered products, returns from
$t_f-i\epsilon$ to $0-2i\epsilon$.  Finally, the segment $a=3$, which provides
a path-integral representation of the initial equilibrium density operator at
$t=0$, runs from $0-2i\epsilon$ to $0-i\beta$, where $\beta$ is the inverse of
the initial temperature.  The action $\bar{S}$ appearing in $Z(j_a)$ is
\begin{eqnarray}
\bar{S}(\phi_a)&=&\int d^3x\left[\int_0^{t_f}dt\,{\cal L}(\phi_1)
-\int_0^{t_f}dt{\cal L}(\phi_2)\right.\nonumber\\
&&\qquad\qquad\qquad+\left.i\int_0^{\beta}d\tau\,{\cal L}_E(\phi_3)\right]\,,
\end{eqnarray}
where ${\cal L}(\phi)$ is the original Lagrangian density (in our case, that
given in (\ref{action})), while ${\cal L}_E$ is the Euclidean version associated
with the density operator, namely
\begin{equation}
{\cal L}_E(\phi_3)=\frac{1}{2}\left(\partial_{\tau}\phi_3\right)^2
+\frac{1}{2}\left(\nabla\phi_3\right)^2-\frac{1}{2}m^2(0)\phi_3^2
+\frac{\lambda}{4!}\phi_3^4\,.
\end{equation}
The source term in (\ref{generatingfunctional}) is
\begin{eqnarray}
\int d^4x\,j\cdot\phi&\,&= \nonumber\\
\int d^3x&&\left\{\int_0^{t_f}dt\,\left[j_1(\bbox{x},t)\phi_1(\bbox{x},t)
+j_2(\bbox{x},t)\phi_2(\bbox{x},t)\right]\right.\nonumber\\
&&\qquad\qquad\quad\qquad
+\left.\int_0^{\beta}d\tau\,j_3(\bbox{x},\tau)\phi_3(\bbox{x},\tau)\right\}\,.
\nonumber\\&&
\end{eqnarray}
Of particular importance are the real-time 2-point functions
($\alpha,\beta=1,2$) given by
\begin{eqnarray}
G_{\alpha\beta}(x,x')&=&-\left.\frac{\partial}{\partial j_{\alpha}(x)}
\frac{\partial}{\partial j_{\beta}(x')}\ln Z(j_a)\right\vert_{j_a=0}\nonumber\\
&=&\pmatrix{
{\rm Tr}[\rho T(\hat{\phi}(x)\hat{\phi}(x'))]&
{\rm Tr}[\rho\hat{\phi}(x')\hat{\phi}(x)]\cr
{\rm Tr}[\rho\hat{\phi}(x)\hat{\phi}(x')]&
{\rm Tr}[\rho\bar{T}(\hat{\phi}(x)\hat{\phi}(x'))]}
\nonumber\\&&
\label{full2ptfunctions}
\end{eqnarray}
where $T$ and $\bar{T}$ denote respectively time-ordered and anti-time-ordered
products of the quantum field operator $\hat{\phi}(x)$ and $\rho$ is the initial
density operator.  Other expectation values can, of course, be obtained from
appropriate derivatives of $Z(j_a)$.

To evaluate these expectation values perturbatively, one splits the action into
an unperturbed part $\bar{S}_0(\phi_a)$, which is quadratic in $\phi_a$, and an
interaction $\bar{S}_{\rm int}(\phi_a)$:
\begin{equation}
\bar{S}(\phi_a)=\bar{S}_0(\phi_a)+\bar{S}_{\rm int}(\phi_a)\,.
\end{equation}
After an integration by parts, $\bar{S}_0$ can be written in terms of a
differential operator ${\cal D}_{ab}$ as
\begin{equation}
\bar{S}_0(\phi_a)
=-\frac{1}{2}\int d^4x\,\phi_a(x){\cal D}_{ab}(x,\partial_\mu)\phi_b(x)\,.
\end{equation}
The lowest-order approximations to the 2-point functions are then the
propagators $g_{ab}(x,x')$, which are solutions of
\begin{eqnarray}
{\cal D}_{ab}(x,\partial_\mu)g_{bc}(x,x')
&=&g_{ab}(x,x'){\cal D}_{bc}(x',-\overleftarrow{\partial}_{\mu'})\nonumber\\
&=&-i\delta_{ac}\delta(x-x')\,,
\label{dgeqdelta}
\end{eqnarray}
subject to appropriate boundary conditions, which arise from continuity of the
$\phi_a$ and their time derivatives around the time path, including the
periodicity condition $\phi_3(\bbox{x},\beta)=\phi_1(\bbox{x},0)$.  The
perturbation series for an expectation value of some product of fields can be
represented in terms of the usual Feynman diagrams, in which propagator lines
representing $g_{ab}$ connect vertices arising from $\bar{S}_{\rm int}$. Since
we consider only a spatially homogeneous system, we shall normally deal with
Fourier transformed Green's functions
\begin{equation}
G_{ab}(t,t';k)=\int d^3(x-x')\,e^{-i\bbox{k}\cdot(\bbox{x}-\bbox{x}')}
G_{ab}(x,x')
\end{equation}
and with propagators $g_{ab}(t,t';k)$ defined in the same way.

The standard choice for $\bar{S}_0$ is simply the quadratic part of $\bar{S}$.
With this choice (and supposing, temporarily, that $m^2(t)<0$, so that there is
no symmetry breaking), one finds that the $g_{ab}(t,t';k)$ are composed of mode
functions $f_k(t)$, which are essentially single-particle wavefunctions,
together with constants which can be interpreted as the occupation numbers $n_k$
of the corresponding single-particle modes.  Because this perturbation theory
has as its lowest-order approximation the theory of a gas of free particles,
which do not scatter, the occupation numbers remain fixed at their initial
values\footnote{In an expanding universe, even a free quantum field theory
reputedly exhibits a phenomenon described as particle creation (see e.g.
\cite{parker1968,birrell1982}).  This arises if, for example, one adopts a
family, parametrized by a time $\bar{t}$, of mode expansions of the
field, with mode functions $f_k^{(\bar{t})}(t)$ which behave approximately as
$\exp[-i\omega_k(t-\bar{t})]$ when $t$ is near $\bar{t}$.  Each choice of a set
of modes in general requires a different set of occupation numbers
$n_k^{(\bar{t})}$ to describe the same physical state, and these may correspond
roughly to the numbers of particles detected by a comoving observer at time
$\bar{t}$.  However, for a given choice of $\bar{t}$, $n_k^{(\bar{t})}$ is
fixed, and does not evolve with time $t$.}.  In the problem at hand, we deal
with an interacting system, driven away from equilibrium by a time-dependent
Hamiltonian and, under these circumstances, one would expect the state of the
system to evolve in response to its changing environment.  In standard
perturbation theory, this state is represented at lowest order by the fixed
occupation numbers.  One therefore suspects that low-order calculations of
time-dependent expectation values (which are all that one can realistically hope
to obtain) will be adequate only over a period of time which is short compared
with some characteristic relaxation time.  The relaxation effects which would
cause occupation numbers (or some appropriate generalization of these) to evolve
with time in the expected manner arise from the absorptive parts of higher-order
loop diagrams, so the perturbation expansion is improved if these contributions
can be partially resummed so as to appear in the propagators $g_{ab}(t,t';k)$.

As explained in detail in \cite{lawrie1988,lawrie1989,lawrie1992}, this
resummation can be achieved by a somewhat more sophisticated choice of the
real-time part of $\bar{S}_0$, namely
\begin{eqnarray}
\bar{S}_0(\phi_1&,&\phi_2) = \bar{S}_0^{(2)}(\phi_1,\phi_2)\nonumber\\
+&&\frac{1}{2}\int_0^{t_f}dt\int\frac{d^3k}{(2\pi)^3}\phi_{\alpha}(k,t)
{\cal M}_{\alpha\beta}(k,t,\partial_t)\phi_{\beta}(-k,t)\,,\nonumber\\
&&\label{S0}
\end{eqnarray}
where $\bar{S}^{(2)}$ is the quadratic part of $\bar{S}$, $\phi_{\alpha}(k,t)$
is the spatial Fourier transform of $\phi_{\alpha}(x)$ and
${\cal M}_{\alpha\beta}$ is a differential operator chosen in the following way.
The propagators $g_{\alpha\beta}(t,t';k)$ now obey the spatial Fourier transform
of (\ref{dgeqdelta}) with a differential operator ${\cal D}_{\alpha\beta}$ that
contains ${\cal M}_{\alpha\beta}$, and we would like them to mimic the full
2-point functions as nearly as possible.  From the hermiticity of $\hat{\phi}$
and of the density operator $\rho$, it is straightforward to show that all the
real-time 2-point functions can be written in terms of a single complex function
$H(x,x')$ as
\begin{equation}
G_{\alpha\beta}(x,x')=H_{\beta}(x,x')\theta(t-t')
+H_{\alpha}(x',x)\theta(t'-t)\,,
\end{equation}
where $H_1(x,x')=H(x,x')$ and $H_2(x,x')=H^*(x,x')$. Clearly, we want
$g_{\alpha\beta}$ to have the same form, and the most general choice of
${\cal D}_{\alpha\beta}$ for which (\ref{dgeqdelta}) admits such a solution is
\begin{equation}
{\cal D}_{\alpha\beta}=\pmatrix{\partial_t^2+\beta_k-i\alpha_k&
\gamma_k\partial_t+\textstyle{\frac{1}{2}}\dot{\gamma}_k+i\alpha_k\cr
-\gamma_k\partial_t-\textstyle{\frac{1}{2}}\dot{\gamma}_k+i\alpha_k&
-\partial_t^2-\beta_k-i\alpha_k\cr}\,,
\label{diffop}
\end{equation}
where $\alpha_k(t)$, $\beta_k(t)$ and $\gamma_k(t)$ are undetermined, real
functions.  These functions can be determined in a self-consistent manner
through an appropriate renormalization prescription. Thus, with the choice
(\ref{S0}) for $\bar{S}_0$, the interaction $\bar{S}_{\rm int}$ contains a
counterterm $-\frac{1}{2}\int\phi{\cal M}\phi$ which we require to cancel some
part of the higher-order contributions to $G_{\alpha\beta}$, thereby optimising
$g_{\alpha\beta}$ as an approximation to $G_{\alpha\beta}$.  In this work we
will, in particular, use a renormalization prescription such that $\beta_k(t)$
has the form $\beta_k(t) =k^2+M^2(t)$, corresponding simply to a mass
renormalization, though more general prescriptions are possible in principle.
It is perhaps worth emphasising that the complete action $\bar{S}$ is, of
course, independent of ${\cal M}_{\alpha\beta}$.  The introduction of this
counterterm does not change the full theory, but serves to optimise our choice
of a lowest-order approximation.  The structure of ${\cal M}_{\alpha\beta}$ is
analogous to that of the effective action obtained, for example, in
\cite{hu1994} by integrating out extra environmental degrees of freedom.  Here,
nothing is integrated out, but we might say that each field mode is treated
self-consistently as interacting with an environment provided by the remaining
modes.

With ${\cal D}_{\alpha\beta}$ given by (\ref{diffop}), the solution to
(\ref{dgeqdelta}) can be written as
\begin{equation}
g_{\alpha\beta}(t,t';k)=h_{\beta}(t,t';k)\theta(t-t')
+ h_{\alpha}(t',t;k)\theta(t'-t)\,,
\label{galphabeta}
\end{equation}
where $h_1(t,t';k)=h(t,t';k)$ and $h_2(t,t';k)=h^*(t,t';k)$, with
\begin{eqnarray}
h&&(t,t';k)={\textstyle{1\over 2}}\exp\left(-{\textstyle{{1\over 2}
\int_{t'}^tdt''\gamma_k(t'')}}\right)\nonumber\\
&&\times\left[\left(N_k(t')+1\right)f_k(t)f_k^*(t')
+\left(N_k^*(t')-1\right)f_k^*(t)f_k(t')\right]\,.\nonumber\\
\label{littleh}
\end{eqnarray}
The mode function $f_k(t)$ is a complex solution of
\begin{equation}
\left[\partial_t^2+\beta_k(t)
-\textstyle{1\over 4}\gamma_k^2(t)\right]f_k(t)=0\,,
\label{modeequation}
\end{equation}
satisfying the Wronskian condition
\begin{equation}
\dot{f}_k(t)f_k^*(t)-f_k(t)\dot{f}_k^*(t)=-i\,,
\label{wronskian}
\end{equation}
while
$N_k(t)$ obeys
\begin{eqnarray}
\left[\partial_t+2i\Omega_k(t)-\frac{\dot{\Omega}_k(t)}{\Omega_k(t)}
+\gamma_k(t)\right]
\left[\partial_t\right.&+&\left.\gamma_k(t)\right]N_k(t)\nonumber\\
&&=2i\alpha_k(t)\,,
\label{nequation}
\end{eqnarray}
where $\Omega_k(t)=1/2f_k(t)f_k^*(t)$.  This function is also required to satify
the subsidiary condition
\begin{eqnarray}
\left[\dot{N}_k(t)+\dot{N}_k^*(t)\right]&+&
2i\Omega_k(t)\left[N_k(t)-N_k^*(t)\right]
\nonumber\\
&+&\gamma_k(t)\left[N_k(t)+N_k^*(t)\right]=0\,,
\label{subsnequation}
\end{eqnarray}
in order that
$\partial_t\left.\left[h_k(t,t')-h_k(t',t)\right]\right\vert_{t=t'}=-i$, which
reflects the canonical commutation relation.  This condition is preserved by
(\ref{nequation}), so it can be regarded as a constraint on the initital values
of $N_k$ and $\dot{N}_k$.  These initial values are determined by continuity
conditions around the closed time path, as discussed in detail in
\cite{lawrie1992}.  Qualitatively, we see that $\gamma_k(t)$ is a damping rate
for unstable quasiparticle excitations and that $N_k(t)$ has a loose
interpretation in terms of time-dependent occupation numbers for quasiparticle
modes.  Indeed, with approximations appropriate to very slow evolution, equation
(\ref{nequation}) reduces to a Boltzmann equation \cite{lawrie1989} in which
$\gamma_k(t)$ and $\alpha_k(t)$ provide the standard scattering integral.  In
this work we consider only a real scalar field, but the dissipative formalism
outlined here can be extended to complex scalar fields \cite{lawrie1997} and
(with somewhat greater difficulty) to spin-${1\over 2}$ fermions
\cite{mckernan1998}.  The dressed propagators obtained by somewhat different
methods in \cite{gleiser1994,berera1998} to describe dissipation in systems
close to equilibrium (or, at least, to a steady state) seem to be a special case
of those given here.

For future use, we note that, associated with ${\cal D}_{\alpha\beta}$ is an
operator
\begin{equation}
\tensor{d}_{\alpha\beta}(t,k)=\pmatrix{\tensor{\partial}_t&\gamma_k(t)
\cr-\gamma_k(t)&-\tensor{\partial}_t\cr}\,,
\label{dlr}
\end{equation}
which has the property
\begin{eqnarray}
g(t,t'';k)\tensor{d}(t'',k)&&g(t'',t';k)\nonumber\\
&&=\left\{\matrix{
-ig(t,t';k)\,, & t>t''>t'\cr ig(t,t';k)\,, &  t'>t''>t\cr 0\,, &
{\rm otherwise}\,.}\right.\label{dleftright}
\end{eqnarray}

\subsection{Spontaneously unbroken symmetry\label{sus}}

As indicated above, we wish to follow the progress of a phase transition
starting from a high-temperature state in which the symmetry
$\phi\leftrightarrow -\phi$ is unbroken and so $\langle\phi(x)\rangle=0$ at each
point in space.  Since the dynamics is governed by a Hamiltonian which respects
this symmetry, it is inevitable that $\langle\phi(x)\rangle\equiv 0$ at all
subsequent times.  We nevertheless expect to encounter states in which the
phenomena conventionally associated with spontaneously broken symmetry (that is,
with a non-zero value of $\langle\phi\rangle$) are realised, and will refer to
such states as having ``spontaneously unbroken symmetry''.  Here, we review the
means by which a vacuum state of this kind in Minkowski spacetime can be treated
\cite{lawrie1988b}.  How such a state may be seen to emerge from a phase
transition is the central problem to be addressed in this paper and will be
discussed in detail later on.

Heuristically, we can envisage two candidate vacua, say $\vert+\rangle$ in which
$\langle\phi\rangle=+\sigma/\sqrt{\lambda}$ and $\vert-\rangle$ in which
$\langle\phi\rangle=-\sigma/\sqrt{\lambda}$.  The overlap
$\langle+\vert-\rangle$ of these states vanishes exponentially as the volume of
space becomes infinite, so $\langle+\vert{\cal O}\vert-\rangle$ also vanishes if
${\cal O}$ is any local operator or the integral over all space of a local
operator.  Consequently, if we consider a superposition
$\vert\alpha\rangle =\sqrt{\alpha}\vert+\rangle+\sqrt{1-\alpha}\vert-\rangle$,
for which
$\langle\alpha\vert\phi\vert\alpha\rangle=(2\alpha-1)\sigma/\sqrt{\lambda}$,
then
$\langle\alpha\vert{\cal O}\vert\alpha\rangle=\alpha\langle
+\vert{\cal O}\vert+\rangle +(1-\alpha)\langle-\vert{\cal O}\vert-\rangle$.  If
the symmetry is exact in Nature, then any experimental probe can couple only to
a symmetrical operator, for which
$\langle+\vert{\cal O}\vert+\rangle =\langle-\vert{\cal O}\vert-\rangle$.  In
that case, $\langle\alpha\vert{\cal O}\vert\alpha\rangle$ is independent of
$\alpha$, and no experiment can determine the value of $\langle\phi\rangle$. In
statistical mechanics, we need incoherent sums over states such as
$\vert+\rangle$ and $\vert-\rangle$, but the same principle applies.

At the formal level, we need a means of calculating expectation values of
symmetrical operators without assuming a non-zero expectation value for $\phi$.
In particular, we need a lowest-order Hamiltonian whose eigenstates yield the
Fock space built on $\vert\alpha\rangle$ rather than on $\vert+\rangle$ or
$\vert-\rangle$.  To this end, we take the state of spontaneously unbroken
symmetry to be characterised by the expectation value of $\phi^2$, which is not
constrained by symmetry.  Instead of the conventional field variable $\psi(x)$,
defined by
\begin{equation}
\phi(x)=\sigma/\sqrt{\lambda} + \psi(x)\,,
\label{psidef}
\end{equation}
we deal with a field $\zeta(x)$ defined by
\begin{equation}
\phi^2(x)=U^2+2U\zeta(x)\,,
\label{zetadef}
\end{equation}
where $U^2=\langle\phi^2(x)\rangle$.  Taking $m^2(t)=m_0^2$ in (\ref{action}),
we find
\begin{eqnarray}
{\cal L}=\frac{1}{2}&&\left(1+2U^{-1}\zeta\right)^{-1}\partial_{\mu}
\zeta\partial^{\mu}\zeta-\frac{\lambda}{6}U^2\zeta^2\nonumber\\
&& +U\left(m_0^2-\frac{\lambda}{6}U^2\right)\zeta
+\frac{i}{2}\delta^4(0)\ln\left(1+2U^{-1}\zeta\right)\,,\nonumber\\
\label{zetalagrangian}
\end{eqnarray}
where the last term comes from the functional Jacobian of the transformation and
an irrelevant constant has been dropped.  Since $\langle\zeta\rangle=0$, the
linear term must vanish to leading order, so we identify
$U=\sqrt{6m_0^2/\lambda}(1+O(\lambda))$.  The range of $\zeta$ in the path
integral is, of course, $-U/2\cdots\infty$. But, since $U$ is of order
$\lambda^{-1/2}$, the lower limit can be extended to $-\infty$ at the expense of
corrections of order $e^{-1/\lambda}$, which do not contribute to perturbation
theory. On expanding (\ref{zetalagrangian}) in powers of $\lambda$, we find
\begin{equation}
{\cal L}=\frac{1}{2}\partial_{\mu}\zeta\partial^{\mu}\zeta
- \frac{1}{2}(2m_0^2)\zeta^2+\cdots\,.
\end{equation}
At leading order, we see that particles created by $\zeta$ from the
spontaneously unbroken vacuum have the same mass $\sqrt{2}m_0$ as those created
from the broken-symmetry vacuum by $\psi$, and it is not hard to convince
oneself \cite{lawrie1988b} that these two types of particle are completely
indistinguishable.  In this formulation, interactions arise from the kinetic
term, and there are an infinite number of vertices (though only a finite number
of these appear at a given order in $\lambda$).  Renormalizability follows from
the fact that $\phi^2(x)$ is a multiplicatively renormalizable composite
operator in the standard formulation, and the $\delta^4(0)$ in the Jacobian
serves to cancel quartic divergences arising from the derivative interactions.

It will be of some importance in what follows that the above manipulations do
not, in themselves, assume that $\langle\phi\rangle=0$, but rather leave
$\langle\phi\rangle$ undetermined. We may therefore consider the case that
$\langle\phi\rangle=\sigma/\sqrt{\lambda}$ and $\zeta$ is related to $\psi$ by
\begin{equation}
\zeta(x) = \frac{1}{2U}\left[\sigma^2/\lambda-U^2+2\left(\sigma/\sqrt{\lambda}
\right)\psi(x)+\psi^2(x)\right]\,.
\end{equation}
In particular, we can express the connected two-point function
$\langle\zeta(x)\zeta(y)\rangle_c$ as
\begin{eqnarray}
\langle\zeta(x)\zeta(y)\rangle_c&&
=\frac{\sigma^2}{\lambda U^2}\langle\psi(x)\psi(y)\rangle_c
+\frac{\sigma}{2\sqrt{\lambda}U^2}\langle\psi(x)\psi^2(y)\rangle_c\nonumber\\
&&+\frac{\sigma}{2\sqrt{\lambda}U^2}\langle\psi^2(x)\psi(y)\rangle_c
+\frac{1}{4U^2}\langle\psi^2(x)\psi^2(y)\rangle_c\,.\nonumber\\
&&\label{zetaandpsi}
\end{eqnarray}
In the Minkowski-space theory, one can easily calculate $U$ in terms of $\sigma$
and verify this relation order by order in $\lambda$.  In the nonequilibrium
state resulting from a phase transition that we plan to investigate, the field
$\psi$ is not defined.  Nevertheless, it will be possible to identify by eye
terms corresponding to the various correlators on the right of
(\ref{zetaandpsi}). 

In what follows, we shall consider only the theory of one real scalar field.  It
is appropriate to mention, however, that the analysis of spontaneously unbroken
symmetry outlined here can be extended to the Higgs sector of a spontaneously
broken gauge theory \cite{lawrie1991}.  In accordance with Elitzur's theorem
\cite{elitzur1975}, non-zero expectation values need be assigned only to
operators which are invariant under the gauge and global symmetries of the
theory.  The generation of fermion masses can also be accomplished in this
way.  Consider, for example, a pair of massless fermions $\psi_L$ and $\psi_R$
and a complex scalar $\phi$ with Lagrangian density
\begin{eqnarray}
{\cal L}=\partial_{\mu}\phi^*\partial^{\mu}\phi&+&
\bar{\psi}_Li\dnot\partial\psi_L
+\bar{\psi}_Ri\dnot\partial\psi_R\nonumber\\
&-&f\left(\bar{\psi}_L\psi_R\phi+\bar{\psi}_R\psi_L\phi^*
\right)+\cdots\ .
\end{eqnarray}
With $\langle\phi^*\phi\rangle=U^2$, we write
$\phi(x)=\sqrt{U^2+2U\zeta(x)}e^{i\theta(x)}$ and construct the single massive
spinor
\begin{equation}
\psi(x)=e^{-i\theta(x)/2}\psi_L(x)+e^{i\theta(x)/2}\psi_R(x)
\end{equation}
to obtain
\begin{eqnarray}
{\cal L}=\partial_{\mu}\zeta\partial^{\mu}\zeta&+&
\bar{\psi}(i\dnot\partial-fU)\psi
+(U^2+2U\zeta)\partial_{\mu}\theta\partial^{\mu}\theta\nonumber\\
&+&\bar{\psi}\gamma^{\mu}\gamma^5\psi\partial_{\mu}\theta+\cdots\ .
\end{eqnarray}
Since only the derivatives of $\theta$ appear, we need not assign $\theta$ an
expectation value, and the fermion mass is $f\sqrt{\langle\phi^*\phi\rangle}$.
This indicates both that our formalism can in principle be extended to more
realistic particle-physics models and also that our earlier remarks concerning
the unobservability of $\langle\phi\rangle$ are not invalidated by the linear
Yukawa coupling of $\phi$ to fermions.

\section{Derivation of the Feynman rules\label{feynmanrules}}

The central purpose of this paper is to derive a set of Feynman rules which will
permit us to follow the evolution of the state of our system as the symmetry
becomes spontaneously unbroken, in the sense described in the last section.  As
an intuitive guide to the considerations involved, we offer in Figures 1a and 1b
an artist's impression of a time-dependent effective potential which might be
thought to govern the evolution of $\phi$. At times earlier than, say, $t_1$, it
has a single minimum at $\phi=0$ while at times later than $t_1$ it has two
symmetrically placed minima at $\phi=\pm\phi_0(t)$.  This is no more than an
intuitive guide, because we have given no precise definition of
$V_{\rm eff}(\phi)$.  There are several conventional definitions of effective
potentials, all of which refer to equilibrium situations, and do not necessarily
fit the dynamical role which is often forced upon them.

Figures 1c and 1d show an artist's impression of the probability density
${\cal P}(\phi,t)$ for the field at the spatial point $\bbox{x}$ to have the
value $\phi$ at time $t$.  Because of spatial homogeneity, this probability is
independent of $\bbox{x}$.  Again, this gives only an intuitive expectation.
Perturbation theory gives no ready access to ${\cal P}(\phi,t)$ and its
structure might well be more complicated than that sketched in the figure (and,
of course, we are ignoring the problems of renormalization that would be
encountered in trying to construct a well-defined ${\cal P}(\phi,t)\,$).
Also, ${\cal P}(\phi,t)$ does not by any means give a complete characterization
of the state of the system.  For that, one would need the vastly more
complicated probability density for field configurations over all space.  The
intuition illustrated is that before some time $t_0$, the most likely value of
$\phi$ is zero, whereas after $t_0$, a broken-symmetry state emerges in which
the most likely values are $\pm\sigma(t)/\sqrt{\lambda}$.

These heuristic considerations do not provide a formal basis for the
calculational scheme we wish to propose.  The foregoing discussion is intended
to motivate the assumption we do make, namely that there is some time $t_0$
before which the most likely field values are near zero, and the appropriate
field variable for perturbative calculations is $\phi$, while after $t_0$
the most likely values are near $\pm\sigma(t)/\sqrt{\lambda}$ and the
appropriate field variable is $\zeta$. For brevity, we shall refer to these two
varieties of perturbation theory as the ``$\phi$-theory'' and the
``$\zeta$-theory''. This situation presents two difficulties.  One is that in
the $\phi$-theory the perturbative expansion of $\langle\phi^2\rangle$ has a
leading term of order $\lambda^0$, while in the $\zeta$-theory the leading term
is of order $\lambda^{-1}$. Thus, although both perturbation series are
expansions in powers of $\lambda$, they are not the same expansion.  We will
propose a concrete means of dealing with this later on.  The second difficulty
is that we need to make the change of variable (\ref{zetadef}) only for times
later than $t_0$, and thus to evaluate a path integral of the form
\begin{equation}
\int_{t<t_0}[d\phi]\int_{t>t_0}[d\zeta]e^{i\bar{S}(\phi,\zeta)}\,.
\end{equation}
This does not factorise into two independent integrals, on account of the
boundary condition that $\phi^2(t_0)=U^2(t_0)+2U(t_0)\zeta(t_0)$.  Indeed, the
nonequilibrium state of $\zeta$ at time $t_0$ is determined by the evolution of
$\phi$ between $t=0$ and $t=t_0$, and must be incorporated through a correct
handling of the boundary conditions at $t_0$.  The evaluation of this path
integral is a hazardous undertaking, and the method we propose is not entirely
rigorous.  We shall encounter various ill-defined quantities, and sensible
prescriptions for dealing with these will be needed.  We therefore begin by
explaining our strategy in the context of a toy model, for which we obtain what
are manifestly the right answers.


\subsection{A toy path integral\label{toy}}

The toy model in question is simply a free scalar field theory in Minkowski
spacetime.  The real-time part of the action (including a source
$J_{\alpha}(t)$) is
\begin{eqnarray}
\bar{S}=\int_0^{t_f}dt&&\left[{\textstyle{1\over 2}}\left(\dot{\phi}_1^2(t)
-\dot{\phi}_2^2(t)\right)\right.\nonumber\\
&&\quad+\left.\left(J_1(t)\phi_1(t)+J_2(t)\phi_2(t)\vphantom{\dot{\phi}}
\right)\right]\,.
\label{toyaction}
\end{eqnarray}
For notational clarity, we do not indicate explicitly the spatial integral or
the gradient and mass terms, which play no direct role in this part of our
analysis. We will evaluate the generating functional given by standard methods
as
\begin{eqnarray}
Z(J_{\alpha})&=&\int[d\phi]e^{i\bar{S}(\phi,J)}\nonumber\\
&=&{\rm const}\times\exp\left[-{\textstyle{1\over 2}}
\int_0^{t_f}dtdt'J_{\alpha}(t)g_{\alpha\beta}(t,t')J_{\beta}(t')\right]
\nonumber\\&&
\label{rightanswer}
\end{eqnarray}
by integrating first over $\phi_{\alpha}(t)$ for $t>t_0$ and then independently
over the remaining fields. We write the sources as
$J_{\alpha}(t)=j_{\alpha}(t)\theta(t_0-t)+l_{\alpha}(t)\theta(t-t_0)$ and
in the first instance set $j_{\alpha}(t)=0$. The propagator
$g_{\alpha\beta}(t,t')$ now obeys (\ref{dgeqdelta}) in the form
$\epsilon_{\alpha\gamma}\partial_t^2g_{\gamma\beta}=
-i\delta_{\alpha\beta}\delta(t-t')$, where $\epsilon_{11}=-\epsilon_{22}=1$ and
$\epsilon_{12}=\epsilon_{21}=0$, and we have again suppressed the spatial
derivative and mass terms.  We define
\begin{equation}
L_{\alpha}(t)=\int_{t_0}^{t_f}dt'g_{\alpha\beta}(t,t')l_{\beta}(t')\,,
\end{equation}
which clearly satisfies $\epsilon_{\alpha\beta}\partial_t^2L_{\beta}(t)
=-il_{\alpha}(t)$, and make the change of integration variable
\begin{equation}
\phi_{\alpha}(t)\to\phi_{\alpha}(t)+iL_{\alpha}(t)\theta(t-t_0)\,.
\label{fieldshift}
\end{equation}
After an integration by parts using the boundary condition
$\phi_1(t_f)=\phi_2(t_f)$, which also entails $l_1(t_f)=l_2(t_f)$, the action
(\ref{toyaction}) becomes
\begin{eqnarray}
\bar{S}\,&=&{\textstyle{1\over 2}}\int_0^{t_f}\epsilon_{\alpha\beta}
\dot{\phi}_{\alpha}(t)\dot{\phi}_{\beta}(t)dt
+\left.i\epsilon_{\alpha\beta}L_{\alpha}(t)
\tensor{\partial}_t\phi_{\beta}(t)\right\vert_{t=t_0}\nonumber\\
&&+{\textstyle{i\over 2}}\int_{t_0}^{t_f}l_{\alpha}(t)g_{\alpha\beta}(t,t')
l_{\beta}(t')dtdt'\nonumber\\
&&-{\textstyle{1\over 2}}\left[L_1^2(t_0)-L_2^2(t_0)\right]\delta(0)\,,
\label{firststep}
\end{eqnarray}
provided that we set $\theta(0)={1\over 2}$.  Here, the fields for $t>t_0$ are
decoupled from the sources, so we can integrate them out.  To do this, we note
that $\exp[i\bar{S}(\phi_1,\phi_2)]$ arises from the product of two time
evolution operators:
\begin{eqnarray}
\int&&[d\phi](\cdots)e^{i\bar{S}(\phi_1,\phi_2)}
={\rm Tr}\left[(\cdots)U_2^{-1}(0,t_f)U_1(0,t_f)\right]\nonumber\\
&&={\rm Tr}\left[(\cdots)U_2^{-1}(0,t_0)U_2^{-1}(t_0,t_f)U_1(t_0,t_f)
U_1(0,t_0)\right],
\nonumber\\&&
\end{eqnarray}
where $U_i$ is the time evolution operator in the presence of the source $J_i$,
and that the derivation of the path integral implies the boundary conditions
$\phi_1(t_f)=\phi_2(t_f)$ and $\dot{\phi}_1(t_f)=\dot{\phi}_2(t_f)$.  In the
absence of sources between $t_0$ and $t_f$, the product
$U_2^{-1}(t_0,t_f)U_1(t_0,t_f)$ is the identity.  Consequently, the path
integration for $t_0<t\le t_f$ yields a factor of 1 together with the boundary
conditions $\phi_1(t_0)=\phi_2(t_0)$ and $\dot{\phi}_1(t_0)=\dot{\phi}_2(t_0)$
on the remaining integral.

Before evaluating the remaining integral, we reinstate a non-zero source $j(t)$
for times before $t_0$.  In the present example, this could have been retained
throughout, at the expense only of additional terms in our equations which were
irrelevant until now.  When dealing with the dynamics of symmetry breaking,
however, we will have more cogent reasons for introducing $j(t)$ only at this
point.  The quantity still to be integrated is now $\exp[i\bar{S}'(\phi,J)]$,
where
\begin{equation}
\bar{S}'=\int_0^{t_0}\left[{\textstyle{1\over 2}}\epsilon_{\alpha\beta}
\dot{\phi}_{\alpha}(t)
\dot{\phi}_{\beta}(t)+\left(j_{\alpha}(t)+\tilde{L}_{\alpha}(t)\right)
\phi_{\alpha}(t)\right]dt\,.
\end{equation}
The quantity $\tilde{L}_{\alpha}(t)$ is a distribution concentrated at $t=t_0$
such that for any pair of test functions $f_{\alpha}(t)$
\begin{equation}
\int_0^{t_0}\tilde{L}_{\alpha}(t)f_{\alpha}(t)
=\left.i\epsilon_{\alpha\beta}L_{\alpha}(t)
\tensor{\partial}_tf_{\beta}(t)\right\vert_{t=t_0}\,.
\label{ltilde}
\end{equation}
Evaluating the integral in the standard way, we obtain
${\rm const}\times\exp[-{1\over 2}{\cal J}]$, with
\begin{eqnarray}
{\cal J}=\int_0^{t_0}\left[j_{\alpha}(t)+\vphantom{\tilde{L}}\right.&&\left.
\tilde{L}_{\alpha}(t)\right]\bar{g}_{\alpha\beta}(t,t')\nonumber\\
&&\times\left[j_{\beta}(t')+\tilde{L}_{\beta}(t')\right]dtdt'\,.
\end{eqnarray}
The propagator $\bar{g}_{\alpha\beta}(t,t')$ is a solution of the same equation
as $g_{\alpha\beta}(t,t')$.  If we assume that these are the same solutions
(which is true if they and their first derivatives coincide when $t=t_0$ or
$t'=t_0$), then we can use (\ref{dleftright}) in the form
\begin{eqnarray}
g_{\alpha\gamma}&&(t,t'')\epsilon_{\gamma\delta}\tensor{\partial}_{t''}
g_{\delta\beta}(t'',t)=-ig_{\alpha\beta}(t,t')\nonumber\\
&&\times\left[\theta(t-t'')\theta(t''-t')-\theta(t'-t'')\theta(t''-t)\right]
\label{gdgeqg}
\end{eqnarray}
to find
\begin{eqnarray}
\int_0^{t_0}dtdt'&&\tilde{L}_{\alpha}(t)g_{\alpha\beta}(t,t')j_{\beta}(t')
\nonumber\\
&&=\int_{t_0}^{t_f}dt\int_0^{t_0}dt'l_{\alpha}(t)g_{\alpha\beta}(t,t')
j_{\beta}(t')
\end{eqnarray}
and
\begin{eqnarray}
\int_0^{t_0}dtdt'\tilde{L}_{\alpha}(t)g_{\alpha\beta}(t,&&t')
\tilde{L}_{\beta}(t')\nonumber\\
&&=-i\left[L_1^2(t_0)-L_2(t_0)\right]\delta(0)\,.
\label{canceldelta}
\end{eqnarray}
Combining these results with those in (\ref{firststep}), we recover the expected
result (\ref{rightanswer}).  In particular, the terms proportional to
$\delta(0)$ cancel. It is straightforward to verify that these manipulations
also work if we replace the free field theory (\ref{toyaction}) with the
dissipative approximate theory (\ref{S0}), provided that the operator
$\epsilon_{\alpha\beta}\tensor{\partial}_t$ in (\ref{firststep}) and
(\ref{ltilde}) is replaced by $\tensor{d}_{\alpha\beta}$ as defined in
(\ref{dlr}).


\subsection{Path integral for a symmetry-breaking phase transition
\label{pathintegral}}

We are finally ready to undertake our central piece of analysis, which is to
derive a perturbative means of calculating the generating functional
$Z(j_a,l_{\alpha})$ of Green's functions which involve $\phi_a(\bbox{x},t)$
($a=1,2,3$) for $0<t<t_0$ and $\zeta_{\alpha}(\bbox{x},t)$ ($\alpha=1,2$) for
$t_0<t<t_f$.  The derivation is quite lengthy, and we set it out in several
steps.

\paragraph*{Step 1: The generating functional.} In the first instance, we set
$j_a=0$ and define
\begin{eqnarray}
Z(0,l_{\alpha})
=\int&&[d\phi]\exp\left[i\bar{S}(\phi_a)\vphantom{\int_{t_0}^{t_f}}\right.
\nonumber\\
&&+\left.i\int_{t_0}^{t_f}dt\int d^3x\, l_{\alpha}(\bbox{x},t)
\zeta_{\alpha}(\bbox{x},t)\right]\,,
\end{eqnarray}
where
\[
\zeta_{\alpha}(\bbox{x},t) = [2U(t)]^{-1}\left[\phi_{\alpha}^2(\bbox{x},t)
-U^2(t)\right]
\]
and $U^2(t)=\langle\phi_{\alpha}^2(\bbox{x},t)\rangle$.  This is the exact
expectation value which must, of course, be determined self-consistently by
demanding that $\langle\zeta_{\alpha}(\bbox{x},t)\rangle=0$.  The reason for
setting the source $j_a$ for $\phi_a$ equal to zero is that this source breaks
the exact symmetry on which our manipulations depend. We shall introduce $j_a$
at a later stage, and subsequently discuss the exact meaning of the generating
functional obtained by this route.

\paragraph*{Step 2: A change of variable.} We now change the integration
variables from $\phi_a(\bbox{x},t)$ to $\zeta_a(\bbox{x},t)$ on the {\it whole
closed time path}, so as to avoid difficulties over the time derivatives of
fields at $t=t_0$. The transformation is formally legitimate even for $t<t_0$,
but for these times we shall eventually have to undo the transformation in order
to construct a sensible perturbation theory.  The resulting Lagrangian is rather
complicated, owing to the fact that $U(t)$ is now time dependent, and we give
only the real-time part (depending on $\phi_1$ and $\phi_2$) of which we shall
make explicit use.  For later convenience, we write it as the sum of several
terms:
\begin{equation}
{\cal L}={\cal L}_0 + {\cal L}_{\rm int} + {\cal L}_{\rm ct}^{(1)}
+ {\cal L}_{\rm ct}^{(>1)}
+ {\cal L}_{\rm tdct} + {\cal L}_{\rm td} + {\cal L}_{\rm jac}\,.
\end{equation}
Of these, the first is
\begin{equation}
{\cal L}_0=-{\textstyle{1\over 2}}\zeta_{\alpha}{\cal D}_{\alpha\beta}
\zeta_{\beta}+\epsilon_{\alpha}\frac{d}{dt}\left[{\textstyle{1\over 2}}
\zeta\dot{\zeta}\right]_{\alpha}\,,
\end{equation}
where ${\cal D}_{\alpha\beta}$ has the form shown in (\ref{diffop}), but with
$\beta_k(t)=k^2+M^2(t)$.  Here and below, we deal with the spatial Fourier
transforms of fields, propagators, etc., but will generally not indicate
explicitly their dependence on $k$ or the associated momentum integrals.  The
total derivative term, in which $\epsilon_1=-\epsilon_2=1$, combines with the
$\partial_t^2$ terms in
$-{1\over 2}\zeta_{\alpha}{\cal D}_{\alpha\beta}\zeta_{\beta}$ to produce
${1\over 2}\epsilon_{\alpha\beta}\dot{\zeta}_{\alpha}\dot{\zeta}_{\beta}$.

The second contribution
\begin{eqnarray}
{\cal L}_{\rm int}={\textstyle{1\over 4}}&&\left[\ln\left(1+\frac{2}{U}
\zeta\right)-\frac{2}{U}\zeta\right]_{\alpha}\nonumber\\
&&\times\left[-U{\cal D}_{\alpha\beta}
+\left(\ddot{U}+M^2U\right)\epsilon_{\alpha\beta}\right]\zeta_{\beta}
\label{Lint}
\end{eqnarray}
contains the principal interactions. There are derivative interactions of the
kind already encountered in section \ref{sus}, and these have been expressed in
terms of the operator ${\cal D}_{\alpha\beta}$, so that use can be made of
(\ref{dgeqdelta}) in computing Feynman diagrams.  There are also non-derivative
interactions arising from the time dependence of $U(t)$.  The linear counterterm
\begin{equation}
{\cal L}^{(1)}_{\rm ct}=-\left[\ddot{U}-m^2(t)U+(\lambda/6)U^3\right]
\epsilon_{\alpha}\zeta_{\alpha}
\label{Lct1}
\end{equation}
appears because we have taken $U^2(t)$ to be the exact expectation value of
$\hat{\phi}^2$, which means that $\langle\zeta_{\alpha}\rangle\equiv 0$.  This
requirement will be implemented self-consistently by requiring the counterterm
to cancel all higher-order contributions to $\langle\zeta_{\alpha}\rangle$.
Further counterterms, denoted by
\begin{eqnarray}
{\cal L}_{\rm ct}^{(>1)}
=&&\frac{1}{2U}\left[\ddot{U}+M^2U-\frac{\lambda}{3}U^3\right]
\epsilon_{\alpha\beta}\zeta_{\alpha}\zeta_{\beta}\nonumber\\
&&+\frac{1}{4}U\ln\left(1+\frac{2}{U}\zeta\right)_{\alpha}
\bar{\cal M}^{\zeta}_{\alpha\beta}\zeta_{\beta}\,,
\label{Lctgt1}
\end{eqnarray}
arise from the fact that we used a renormalized mass $M(t)$ and the dissipative
coefficients $\alpha_k(t)$ and $\gamma_k(t)$ in ${\cal L}_0$.  Again, these
quantities are to be determined self-consistently by using the counterterms to
cancel appropriate parts of higher-order contributions to the 2-point functions.
The dissipative counterterm
\begin{equation}
\bar{\cal M}^{\zeta}_{\alpha\beta}=\pmatrix{-i\alpha&\gamma\partial_t
+{\textstyle{1\over 2}}\dot{\gamma}+i\alpha\cr-\gamma\partial_t
-{\textstyle{1\over 2}}\dot{\gamma}+i\alpha&-i\alpha\cr}
\end{equation}
differs from the ${\cal M}_{\alpha\beta}$ of section \ref{dissipation} only
insofar as it excludes $M(t)$, which is treated separately.The two additional
counterterms
\begin{equation}
{\cal L}_{\rm tdc}=X_{\alpha}{\cal D}_{\alpha\beta}\zeta_{\beta}
-\zeta_{\alpha}{\cal D}_{\alpha\beta}X_{\beta} +\frac{d}{dt}\left[Y_{\alpha}
\tensor{d}_{\alpha\beta}\zeta_{\beta}^2\right]\,,
\label{Ltdc}
\end{equation}
where $X_{\alpha}(t)$ and $Y_{\alpha}(t,k)$ are as yet undetermined functions,
are total time derivatives.  They will be used to facilitate the handling of
boundary terms arising from integrations by parts in the evaluation of certain
Feynman diagrams.  More terms of this kind might be needed for calculations at
higher orders than we consider explicitly in this paper. The sum of terms given
so far differs from the original Lagrangian by a total time derivative, namely
\begin{eqnarray}
{\cal L}_{\rm td}=\frac{d}{dt}&&\left\{\frac{1}{4}
\left[\ln\left(1+\frac{2}{U}\zeta\right)-\frac{2}{U}\zeta\right]_{\alpha}\right.
\epsilon_{\alpha\beta}\left(U\dot{\zeta}-\dot{U}\zeta\right)_{\beta}\nonumber\\
&&\left.\vphantom{\frac{2}{U}}+\epsilon_{\alpha}\dot{U}\zeta_{\alpha}
-X_{\alpha}\tensor{d}_{\alpha\beta}\zeta_{\beta}
-Y_{\alpha}\tensor{d}_{\alpha\beta}\zeta_{\beta}^2\right\}\,.
\label{Ltd}
\end{eqnarray}
At this point, ${\cal L}_{\rm td}$ could be integrated round the whole time
contour to yield zero, when use is made of the boundary conditions at $t=0$
and $t=t_f$.  We do not do this, because we want to insert an extra boundary at
$t_0$ and to indicate explicitly how we treat the boundary conditions there.
Finally, the Jacobian of the transformation is provided by
\begin{equation}
{\cal L}_{\rm jac}=\frac{i}{2}\delta^4(0)\left[\ln\left(1+\frac{2}{U}\zeta_1\right)
+\ln\left(1+\frac{2}{U}\zeta_2\right)\right].
\label{Ljac}
\end{equation}

\paragraph*{Step 3:  The path integral for $t_0<t<t_f$.}
Our next task is to evaluate the $\zeta$ integral for times after $t_0$.  As
usual, it is necessary to extract the interactions and counterterms as
derivatives with respect to the source. Thus, we write
\begin{equation}
Z(0,l_{\alpha})
=\exp\left[iV^{\zeta}(-i\delta/\delta l_{\alpha})\right]Z_1(l_{\alpha})\,,
\end{equation}
where $V^{\zeta}=\int_{t_0}^{t_f}\left({\cal L}-{\cal L}_0\right)dt$ and
\begin{equation}
Z_1(l_{\alpha})=\int[d\zeta]\exp\left[i\int^{t_0}\,{\cal L}dt
+i\int_{t_0}^{t_f}\left({\cal L}_0+l_{\alpha}\zeta_{\alpha}\right)dt\right]\,.
\end{equation}
The notation $\int^{t_0}dt$ indicates that the integral is over both the
real-time segments of the time path for $0<t<t_0$ and the imaginary time
segment.  We shall assume that the contribution to $V^{\zeta}$ from
${\cal L}_{\rm td}$ can be neglected.  The integral of this contribution over
the closed time path does indeed vanish, provided that the boundary conditions
to be explained in Step 4 below are valid.  We shall later argue on heuristic
grounds that the total derivatives in ${\cal L}_{\rm tdc}$ (which would also
integrate to zero) ought nevertheless to be retained.  Now the manipulations
leading to (\ref{firststep}) can be repeated, with the result
\begin{equation}
Z(0,l_{\alpha})=\exp\left[iV^{\zeta}(-i\delta/\delta l_{\alpha})\right]
\left[\exp\left(-{\textstyle{1\over 2}}{\cal J}^{\zeta}\right)
\int[d\zeta]e^{i\bar{S}_1}\right].
\end{equation}
Here, ${\cal J}^{\zeta}$ is
\begin{eqnarray}
{\cal J}^{\zeta}=\int_{t_0}^{t_f}&& l_{\alpha}(t)g^{\zeta}_{\alpha\beta}(t,t')
l_{\beta}(t')dtdt'
\nonumber\\&&
+i\epsilon_{\alpha\beta}L_{\alpha}(t_0)L_{\beta}(t_0)\delta(0)\,,
\label{jzeta}
\end{eqnarray}
where $g^{\zeta}$ is the propagator for the field $\zeta$, which we shall need
to distinguish from that for $\phi$, and
\begin{equation}
L_{\alpha}(t)=\int_{t_0}^{t_f}\,g^{\zeta}_{\alpha\beta}(t,t')l_{\beta}(t')dt'\,.
\end{equation}
The action in the remaining path integral is
\begin{equation}
\bar{S}_1=\int^{t_0}{\cal L}\,dt +\int_{t_0}^{t_f}{\cal L}_0\,dt
+ i\left.L_{\alpha}(t)
\tensor{d}_{\alpha\beta}(t)\zeta_{\beta}(t)\right\vert_{t=t_0}\,,
\label{S1}
\end{equation}
provided, as we assume, that any effect of the shift analogous to
(\ref{fieldshift}) on the interaction terms in ${\cal L}$ can be neglected.
There is, indeed, no effect if these interaction terms are taken to exist at
all times less than, but not equal to $t_0$, but we are unable to prove that
this is really legitimate. As in the toy calculation, the path integral for
$t_0<t<t_f$ can now be performed, yielding a factor of 1.

\paragraph*{Step 4:  The path integral for $t<t_0$.}
We now transform the path integration variables from $\zeta$ back to $\phi$. In
making this transformation, we assume that the boundary conditions
$\zeta_1(t_0)=\zeta_2(t_0)$ and $\dot{\zeta}_1(t_0)=\dot{\zeta}_2(t_0)$
translate into $\phi_1(t_0)=\phi_2(t_0)$ and
$\dot{\phi}_1(t_0)=\dot{\phi}_2(t_0)$.  Indeed, we have already implicitly
assumed that the same is true at $t_f$.  At the heuristic level, this seem
justified if we regard the path integral as a sum over sufficiently smooth
functions $\phi_{\alpha}(t)$ or $\zeta_{\alpha}(t)$ for which
$d\phi_{\alpha}^2(t)/dt=2\phi_{\alpha}(t)\dot{\phi}_{\alpha}(t)$. The action
in the remaining path integral is just the standard action for $\phi$, except
for the boundary term in (\ref{S1}). At this point, we add sources $j_a(t)$ for
the fields $\phi_a(t)$, thereby defining
\begin{eqnarray}
Z(j_a,l_{\alpha})
=\exp&&\left[iV^{\zeta}(-i\delta/\delta l_{\alpha})\right]\nonumber\\
&&\times\left[\exp\left(-{\textstyle{1\over 2}}{\cal J}^{\zeta}\right)
\int_{t<t_0}[d\phi]e^{i\bar{S}_2}\right]\,,
\end{eqnarray}
with
\begin{eqnarray}
\bar{S}_2=\int^{t_0}dt\left\{{\cal L}\vphantom{\tilde{L}_{\alpha}}\right.
&& +j_a(t)\phi_{\alpha}(t)\nonumber\\
&&\left. + {\textstyle{i\over 2}}
\tilde{L}_{\alpha}(t)\left[\phi_{\alpha}^2(t)-U^2(t)\right]\right\}\,,
\label{S2}
\end{eqnarray}
where now $\tilde{L}_{\alpha}(t)$ is a distribution such that
\begin{equation}
\int^{t_0}\tilde{L}_{\alpha}(t)f_{\alpha}(t)dt=\left.L_{\beta}(t)
\tensor{d}_{\beta\alpha}(t)
\left(\frac{f_{\alpha}(t)}{U(t)}\right)\right\vert_{t=t_0}\,.
\end{equation}
It should be apparent that $Z(j_a,l_{\alpha})$ has the following significance:
(i) $Z(0,l_{\alpha})$ correctly generates the expectation values we seek for
$t_0<t<t_f$;  (ii) $Z(j_a,0)$ correctly generates expectation values for
$0<t<t_0$, which we also need; (iii) however, because of the way in which $j_a$
has been introduced, $Z(j_a,l_{\alpha})$ does {\it not} yield correctly the
expectation values that would exist after $t_0$ if a real, physical source
$j(t)$ had been present at earlier times. This does not concern us, since for us
$j_a(t)$ is merely a technical device for generating the expectation values of
the source-free theory.

Formally, the remaining path integral can be evaluated by standard methods,
with the result
\begin{eqnarray}
Z(j_a,l_{\alpha})=\exp\left[iV^{\zeta}(-i\delta/\delta l_{\alpha})\right.+&&
\left.iV^{\phi}(-i\delta/\delta j_a)\right]\nonumber\\
&&\times Z_0(j_a,l_{\alpha})\,,
\label{finalgenfunc}
\end{eqnarray}
where $V^{\phi}$ represents the interactions of $\phi$ for $t<t_0$.  Here,
dissipative effects are taken into account by a counterterm
$-{1\over 2}\phi_{\alpha}\bar{\cal M}^{\phi}_{\alpha\beta}\phi_{\beta}$ in which
$\bar{\cal M}^{\phi}_{\alpha\beta}$ has the same structure as
$\bar{\cal M}^{\zeta}_{\alpha\beta}$, but with different coefficients
$\alpha_k(t)$ and $\gamma_k(t)$ associated with the behaviour of the $\phi$
correlator.  The functional $Z_0(j_a,l_{\alpha})$ is conveniently written as
$\exp\left(-{1\over 2}{\cal K}\right)$, with
\begin{equation}
{\cal K}={\cal K}_{\zeta}+{\cal K}_{\phi\zeta}+{\cal K}_{\rm tr}
+{\cal K}_{\delta}+{\cal K}_0\,.
\label{K}
\end{equation}
The various terms arise from different stages of the foregoing calculation as
follows. The first,
\begin{equation}
{\cal K}_{\zeta}=\int_{t_0}^{t_f}l_{\alpha}(t)g^{\zeta}_{\alpha\beta}(t,t')
l_{\beta}(t')dtdt'\,,
\label{Kzeta}
\end{equation}
provides just the $\zeta$ propagator as in (\ref{jzeta}).  The second,
\begin{equation}
{\cal K}_{\phi\zeta}=\int^{t_0}j_a(t)g^{\phi}_{ab}(t,t';l)j_b(t')dtdt'
\label{Kphizeta}
\end{equation}
arises in the same way from the path integral over $\phi$.  However, the
boundary term in (\ref{S2}), being quadratic in $\phi$, leads to a modified
differential operator $\bar{\cal D}^{\phi}_{ab}(t;l)$, whose real-time
components
\begin{equation}
\bar{\cal D}^{\phi}_{\alpha\beta}(t;l)={\cal D}^{\phi}_{\alpha\beta}(t)
-i\pmatrix{\tilde{L}_1(t)&0\cr 0&\tilde{L}_2(t)}\,,
\end{equation}
depend on the source $l_{\alpha}(t)$ for $\zeta(t)$. The propagator
$g^{\phi}_{ab}(t,t';l)$ is a solution of
\begin{equation}
\bar{\cal D}^{\phi}_{ac}(t;l)g^{\phi}_{cb}(t,t';l)=-i\delta_{ab}\delta(t-t')\,,
\label{dbargeqdelta}
\end{equation}
and is therefore also a functional of $l_{\alpha}(t)$.  For this reason too, the
final Gaussian path integral that remains after the extraction of
${\cal K}_{\zeta}$ and ${\cal K}_{\phi\zeta}$ depends on $l_{\alpha}(t)$ and on
evaluating it we find
\begin{equation}
{\cal K}_{\rm tr}={\rm Tr}\,\ln\bar{\cal D}^{\phi}(l)\,.
\label{Ktr}
\end{equation}
Finally,
\begin{equation}
{\cal K}_{\delta}=i\epsilon_{\alpha\beta}L_{\alpha}(t_0)L_{\beta}(t_0)\delta(0)
\label{Kdelta}
\end{equation}
is the last term of (\ref{jzeta}) and
\begin{eqnarray}
{\cal K}_0&=&-\sum_{\alpha} \int^{t_0}dt \tilde{L}_{\alpha}(t)U^2(t)\nonumber\\
&=&-\left.\left[L_1(t)-L_2(t)\right]\left(\tensor{\partial}_t
+\gamma^{\zeta}(t)\right)U(t)
\right\vert_{t=t_0}
\label{K0}
\end{eqnarray}
is the last term of (\ref{S2}).  The superscript on $\gamma^{\zeta}(t)$
indicates that this is the damping rate for $\zeta$ rather than for $\phi$.


\subsection{The Feynman rules\label{rules}}

In the usual way, perturbation theory now consists in expanding the interactions
$V^{\phi}$ and $V^{\zeta}$ in (\ref{finalgenfunc}) in powers of $\lambda$, and
we shall shortly describe the diagrammatic rules through which the terms of the
perturbation series can be represented.  First, however, we must confront a
difficulty which was postponed earlier.  Namely, the perturbation theory for
$t<t_0$, based on the field $\phi$ yields an expectation value for $\phi^2$
which is of order $\lambda^0$, whereas the scheme based on $\zeta$ for $t>t_0$
yields a leading term of order $\lambda^{-1}$.  Indeed, a perturbative treatment
of the $\zeta$-theory is possible only if we take
$U(t)=\sqrt{\langle\phi^2(t)\rangle}=O\left(\lambda^{-1/2}\right)$.

We propose to resolve this conflict in the following way.  For $t>t_0$, we
define $v(t)=\lambda^{1/2}U(t)$.  Then the expectation value of a quantity
$A(\zeta)$ can be estimated in the form
\begin{equation}
\langle A(\zeta)\rangle= \lambda^n\left[a_0(v)+\lambda a_1(v)
+\lambda^2a_2(v)+\cdots\right]\,,
\label{expectA}
\end{equation}
which is a power series in $\lambda$, provided that $v$ is formally regarded
as being of order $\lambda^0$.  For $v(t)$, we will have an equation of motion
roughly of the form
\begin{equation}
\ddot{v}=f_0(v)+\lambda f_1(v)+\lambda^2f_2(v)+\cdots\ ,
\label{eofm}
\end{equation}
whose right hand side is again a formal power series in $\lambda$. The desired
expectation value will then be obtained by substituting into (\ref{expectA}) the
solution of (\ref{eofm}) which satisfies the appropriate initial conditions. The
initial values $v(t_0)$ and $\dot{v}(t_0)$ are calculable in the $\phi$-theory
as power series in $\lambda$, but are of order $\lambda^{1/2}$. In principle,
each of these power series can be truncated at any desired order (or some
partial resummation technique might perhaps be devised). However, the solution
of (\ref{eofm}) with the appropriate initial conditions {\it cannot} be
expressed as a power series in $\lambda$ and nor, therefore, can the final
result for $\langle A(\zeta)\rangle$.  Nevertheless, this scheme leads to a
sequence of approximations which in principle can be systematically pursued to
arbitrary orders. In view of the evident complexity of the nonequilibrium
theory, though, a determination of the convergence or summability properties of
this sequence is well beyond the analytical powers of the present author.

With this approximation scheme in mind, we now describe the Feynman rules for
constructing the required power series.  For $t<t_0$, the rules are just the
standard ones for $\lambda\phi^4$ theory, with propagators constructed according
to the prescription described in section \ref{dissipation}. For $t\ge t_0$,
things are more complicated:  we discuss first the propagators and then the
vertices.


\subsubsection{The propagators\label{propagators}}

In standard perturbation theory, one has a lowest-order generating functional
$Z_0(j_a,l_{\alpha})=\exp[-{1\over 2}{\cal K}]$ in which ${\cal K}$ is a
quadratic functional of the sources, giving rise to propagator lines which
connect vertices, and there is one such propagator for each particle species.
Here, ${\cal K}$ is the nonlinear functional defined by (\ref{K})-(\ref{K0})
and this gives rise to an additional infinite set of propagators which connect
times before and after $t_0$.  In effect, these propagators represent the
nonequilibrium density matrix $\rho(t_0)$ which, in the Schr\"{o}dinger picture,
would result from evolving the initial state from $t=0$ to $t=t_0$. 

Of the contributions to ${\cal K}$ listed in (\ref{K}), the first,
${\cal K}_{\zeta}$, yields just the propagator $g_{\alpha\beta}^{\zeta}(t,t')$,
which is indicated by the solid line of Figure 2(a).  The second contribution,
${\cal K}_{\phi\zeta}$ involves $g^{\phi}_{ab}(t,t';l)$, which is a solution of
(\ref{dbargeqdelta}).  This solution can be written as
\begin{eqnarray}
g^{\phi}_{ab}(t,t';l)=\sum_{n=0}^{\infty}(-1)^n&&
\int dt_1\cdots dt_n\,g^{\phi}_{a\alpha}(t,t_1)
\tilde{L}_{\alpha\beta}(t_1)\nonumber\\
&&\times g^{\phi}_{\beta\gamma}(t_1,t_2)\tilde{L}_{\gamma\delta}(t_2)\cdots
g^{\phi}_{\epsilon b}(t_n,t')\,,\nonumber\\&&
\end{eqnarray}
where $\tilde{L}_{\alpha\beta}$ is the array with $\tilde{L}_{11}=\tilde{L}_1$,
$\tilde{L}_{22}=\tilde{L}_2$ and $\tilde{L}_{12}=\tilde{L}_{21}=0$.  The first
term of this series ($n=0$) is just the $\phi$ propagator $g^{\phi}_{ab}(t,t')$,
depicted by the broken line of Figure 2(b). The term $n=1$ is equal to
$\int_{t_0}^{t_f}g_{ab\gamma}(t,t',t'')l_{\gamma}(t'')dt''$.  The new
propagator 
\begin{eqnarray}
g_{ab\gamma}&&(t,t',t'')\nonumber\\
&&=\left.\sum_{\alpha\beta}\left[\frac{g^{\phi}_{a\alpha}(t,t_1)
g^{\phi}_{\alpha b}(t_1,t')}{U(t_1)}\right]\tensor{d}_{\alpha\beta}(t_1)
g^{\zeta}_{\beta\gamma}(t_1,t'')\right\vert_{t_1=t_0}\nonumber\\&&
\end{eqnarray}
is depicted in Figure 2(c), where the open circle denotes
$U(t_1)^{-1}\tensor{d}(t_1)\vert_{t_1=t_0}$.  The terms $n=2$ and $n=3$ are
shown in Figures 2(d) and 2(e).  The dotted lines represent $g^{\phi}(t_i,t_j)$
in which, after performing the derivatives in $\tensor{d}$, $t_i$ and $t_j$ are
set equal to $t_0$.  Explicit expressions for these propagators are
straightforwardly obtained, but are somewhat cumbersome and will not be
reproduced here.

In the same way, ${\cal K}_{\rm tr}$ defined in (\ref{Ktr}) is given (apart
from an irrelevant constant) by

\vbox{\begin{eqnarray}
{\cal K}_{\rm tr}=-\sum_{n=1}^{\infty}\frac{(-1)^n}{n}\int dt_1&&
\cdots dt_n\,{\rm tr}\left[
\tilde{L}(t_1)g^{\phi}(t_1,t_2)\tilde{L}(t_2)\right.\nonumber\\
&&\left.\times g^{\phi}(t_2,t_3)\tilde{L}(t_3)\cdots
g^{\phi}(t_n,t_1)\right],\nonumber\\&&
\end{eqnarray}}
The propagators corresponding to the first few terms of this series are
depicted in Figures 2(g) - 2(i).  For example, the first term is
$\int_{t_0}^{t_f}g_{\alpha}(t)l_{\alpha}(t)dt$, where
\begin{eqnarray}
g_{\alpha}(t)=-\left[\frac{h^{\phi}(t_1,t_1)}{U(t_1)}\right]&&
\left(\tensor{\partial}_{t_1}-\gamma^{\zeta}(t_1)\right)\nonumber\\
&&\times\left.\left[g_{1\alpha}^{\zeta}(t_1,t)
-g_{2\alpha}^{\zeta}(t_1,t)\right]\right\vert_{t_1=t_0}.\nonumber\\&&
\label{galpha}
\end{eqnarray}
Here, use has been made of the fact that
$g^{\phi}_{\alpha\beta}(t,t)=h^{\phi}(t,t)$ is independent of $\alpha$ and
$\beta$, as is easily checked from (\ref{galphabeta}) and (\ref{littleh}). In
this propagator, the two time arguments of $g^{\phi}(t,t')$ are set equal before
acting with the derivatives in $\tensor{d}$.  Later terms in the series for both
${\cal K}_{\phi\zeta}$ and ${\cal K}_{\rm tr}$ involve derivatives of
$g^{\phi}(t,t')$ whose time arguments are to be set equal after differentiation,
and these equal-time limits are, unfortunately, not unambiguously defined.  They
can, of course, be made well-defined by specifying the order in which limits are
to be taken, but we are not able to offer a well-motivated general prescription
for how this should be done.  We shall, however, describe examples of low-order
calculations in which the appropriate prescription can be ascertained with
reasonable plausibility.

The contribution ${\cal K}_0$ to ${\cal K}$ defined in (\ref{K0}) can be written
as $\int_{t_0}^{t_f}g^0_{\alpha}(t)l_{\alpha}(t)dt$, with
\begin{eqnarray}
g^0_{\alpha}&&(t)\nonumber\\
&&=\left.U(t_1)\left(\tensor{\partial}_{t_1}-\gamma^{\zeta}(t_1)\right)
\left[g^{\zeta}_{1\alpha}(t_1,t)-g^{\zeta}_{2\alpha}(t_1,t)\right]
\right\vert_{t_1=t_0}\,,
\nonumber\\&&
\label{g0alpha}
\end{eqnarray}
and is represented in Figure 2(f).  It evidently has a structure similar to that
of (\ref{galpha}), and the significance of this will become clear below.

Compared with diagrammatic rules of the conventional kind, most of the
propagators shown in Figure 2 look like vertices, so it is perhaps worth
emphasising that they really are propagators. They must be used to connect the
vertices described below in order to form a valid diagram. For example, if the
object in Figure 2(c) were a vertex, it could be combined with the propagator of
Figure 2(b) to form a diagram with the topology of Figure 2(g), but this does
not result in a valid expression. To make this apparent at the visual level, we
envisage the propagators as ``cables'' terminated by ``plugs'' (indicated by the
arrows in Figure 2) and will depict genuine vertices as possessing ``sockets''
that accept such plugs.  Then, a valid diagram can be wired up by plugging
cables into sockets, but not by tying cables together. Formally, plugs
corresponds to the sources $j_a$ or $l_{\alpha}$ which accompany the propagators
in $Z_0(j_a,l_{\alpha})$, while sockets correspond to the derivatives in
$V^{\phi}(-i\delta/\delta j)$ and $V^{\zeta}(-i\delta/\delta l)$.  Once a valid
diagram is formed, account must also be taken of the spatial momenta which we
have not indicated explicitly. These appear in the standard way, being conserved
both at genuine vertices and at the circled vertices internal to the composite
propagators.  Each of these circled vertices contains the operator
$\tensor{d}(t;k)$ whose momentum $k$ is that of the single $\zeta$ propagator
emerging from it.


\subsubsection{The vertices\label{vertices}}

The interactions $V^{\phi}$ and $V^{\zeta}$ in (\ref{finalgenfunc}) are, of
course, represented by vertices.  Interactions of $\phi$ exist only before $t_0$
while those of $\zeta$ exist only after $t_0$ and there are no vertices
involving both fields.  For $\phi$, we introduce a renormalized mass $\mu(t)$
and a dissipative counterterm $\bar{\cal M}^{\phi}_{\alpha\beta}(t)$ as
described in section \ref{dissipation}.  The interaction part of the Lagrangian
density is then
\begin{eqnarray}
{\cal L}^{\phi}_{\rm int}={1\over 2}\left[m^2(t)+\mu^2(t)\right]&&
\phi_{\alpha}\epsilon_{\alpha\beta}\phi_{\beta}
+{1\over 2}\phi_{\alpha}\bar{\cal M}^{\phi}_{\alpha\beta}(t)\phi_{\beta}
\nonumber\\
&&-\frac{\lambda}{4!}\left(\phi_1^4-\phi_2^4\right)\,.
\end{eqnarray}
The vertices corresponding to the quartic interaction, the mass counterterm and
the dissipative counterterm are shown in Figures 3(a), 3(f) and 3(g)
respectively.

Interactions of $\zeta$ are contained in (\ref{Lint}) - (\ref{Ltdc}) and
(\ref{Ljac}).  In order to do perturbation theory, we set
$U(t)=\lambda^{-1/2}v(t)$ and expand in powers of $\lambda$.
For ${\cal L}_{\rm int}$, we obtain
\begin{eqnarray}
{\cal L}^{\zeta}_{\rm int}&=&-{\textstyle{1\over 2}}\lambda^{1/2}
v^{-2}\left(\ddot{v}
+M^2 v\right)\left(\zeta_1^3-\zeta_2^3\right)\nonumber\\
&&+{\textstyle{1\over 2}}\lambda^{1/2}v^{-1}(\zeta^2)_{\alpha}
{\cal D}^{\zeta}_{\alpha\beta}
\zeta_{\beta}\nonumber\\
&&+{\textstyle{2\over 3}}\lambda v^{-3}\left(\ddot{v}+M^2 v\right)
\left(\zeta_1^4-\zeta_2^4\right)\nonumber\\
&&-{\textstyle{2\over 3}}\lambda v^{-2}(\zeta^3)_{\alpha}
{\cal D}^{\zeta}_{\alpha\beta}\zeta_{\beta} +\cdots\ .
\label{lint}
\end{eqnarray}
The corresponding vertices are those shown in Figures 3(b) - 3(e), where a
stroke on one of the legs indicates the action of
${\cal D}^{\zeta}_{\alpha\beta}$ on the propagator attached to that leg.
The linear counterterm (\ref{Lct1}) becomes
\begin{equation}
{\cal L}^{(1)}_{\rm ct}=-\lambda^{-1/2}\left(\ddot{v}-m^2v
+{\textstyle{1\over 6}}v^3\right)\epsilon_{\alpha}\zeta_{\alpha}
\label{linearct}
\end{equation}
and is represented by Figure 3(h).  The counterterms contained in (\ref{Lctgt1})
include a mass renormalization
\begin{equation}
{\cal L}^{(\rm mass)}_{\rm ct}={\textstyle{1\over 2}}v^{-1}\left(\ddot{v}+M^2v-
{\textstyle{1\over 3}}v^3\right)\left(\zeta_1^2-\zeta_2^2\right)\,,
\label{massct}
\end{equation}
depicted in Figure 3(k), of which we shall make explicit use, together with
counterterms associated with $\bar{\cal M}^{\zeta}_{\alpha\beta}$.  The first
two of these are shown in Figures 3(l) and 3(p), but we shall not make explicit
use of them.  Finally, the two total derivatives in (\ref{Ltdc}) are represented
by Figures 3(i) and 3(m), while the first two of the infinite sequence of terms
representing the Jacobian in (\ref{Ljac}) are shown in Figures 3(j) and 3(n).


\section{Equations of motion and continuity conditions at
$\lowercase{t}_0$\label{eomcon}}

We now discuss several calculations which serve the dual purpose of completing
the specification of the Feynman rules derived in the last section and of
illustrating their use.  The Feynman rules are so far incompletely specified for
two reasons.  One is that that they involve the expectation value
$v^2(t)=\lambda\langle\phi^2(t)\rangle$, the renormalized masses $\mu(t)$ and
$M(t)$ and the dissipative coefficients $\alpha^{\phi}_k(t)$,
$\alpha^{\zeta}_k(t)$, $\gamma^{\phi}_k(t)$ and $\gamma^{\zeta}_k(t)$ for which
we have no concrete expressions in hand.  The second is that the propagators
$g^{\phi}_{ab}(t,t';k)$ and $g^{\zeta}_{\alpha\beta}(t,t';k)$ are solutions of
(\ref{dgeqdelta}) (with the appropriate operator ${\cal D}_{ab}$ in each case),
but the appropriate solutions must be identified through suitable boundary
conditions. When these propagators are represented in the form of
(\ref{galphabeta}) and (\ref{littleh}), the remaining ambiguity resides in the
functions $N^{\phi}_k(t)$ and $N^{\zeta}_k(t)$.  These in turn are solutions of
(\ref{nequation}) and it is the initial conditions for these functions that are
needed.

For the unbroken-symmetry state prior to $t_0$, the renormalization and boundary
conditions which determine $\mu(t)$, $\alpha^{\phi}_k(t)$, $\gamma^{\phi}_k(t)$
and $N^{\phi}_k(t)$ are described in \cite{lawrie1989,lawrie1992} and we shall
not repeat the discussion here.  The determination of $\alpha^{\zeta}_k(t)$ and
$\gamma^{\zeta}_k(t)$ presents no new difficulty beyond that of computing the
required integrals and will also not be discussed.  It is therefore the
determination of $v(t)$, $M(t)$ and the initial conditions on $N^{\zeta}_k(t)$
which are principally of interest. Also of concern is the fact that, in
evaluating the path integral for $Z(j_a,l_{\alpha})$, we were unable to treat
the boundary conditions at $t_0$ in a fully rigorous manner, with the result
that certain ambiguities remain.  We will show how these ambiguities can be
resolved at the lowest non-trivial order of our approximation scheme.


\subsection{The condition $\langle\zeta(t)\rangle=0$}

We start with the fact that $U^2(t)=\lambda^{-1}v^2(t)$ was defined to be the
exact expectation value of $\langle \phi^2(t)\rangle$ and is therefore to be
determined self-consistently from the requirement that
$\langle\zeta(t)\rangle=0$.  It is sufficient to ensure that the
one-particle-irreducible contribution $\langle\zeta(t)\rangle_{\rm 1PI}$
vanishes. As illustrated in Figure 4, this quantity can be separated into two
parts,
\begin{equation}
\langle\zeta(t)\rangle_{\rm 1PI}
=\langle\zeta(t)\rangle_{\rm 1PI}^{\rm (anchored)}+
\langle\zeta(t)\rangle_{\rm 1PI}^{\rm (free)}\,.
\end{equation}
Diagrams contributing to the first part contain at least one of the circled
vertices internal to one of the composite propagators, which we will describe
as ``anchoring'' the corresponding time argument at $t=t_0$, whereas all the
vertices in the second part are free to range between $t_0$ and $t_f$. In
Figure 4(a), the diagram labelled (ii) is just the propagator of Figure 2(g),
while that labelled (iii) is constructed from the propagator of Figure 2(c)
together with the ordinary $\phi$ vertex of Figure 3(a).  These are the first
two of a sequence of diagrams whose sum reproduces $\langle\phi^2(t_0)\rangle$
as calculated from the $\phi$-theory.  More precisely, on combining these with
diagram (i) (which arises from Figure 2(f)), we obtain the expression
\begin{eqnarray}
-\frac{i}{2}\left[g^{\zeta}_{\alpha 1}(t,t_1)\right.&-&
\left.g^{\zeta}_{\alpha 2}(t,t_1)\right]
\left(\tensor{\partial}_{t_1}+\gamma^{\zeta}(t_1)\right)\nonumber\\
&&\quad\times\left.\left[U(t_1)-\frac{\langle\phi^2(t_1)\rangle}
{U(t_1)}\right]\right\vert_{t_1=t_0}.
\end{eqnarray}
With the natural requirement that $\langle\phi^2(t)\rangle$ and its first
derivative should be continuous at $t_0$, this vanishes identically.  The fact
that contributions from several different terms of (\ref{K}) conspire to give
this satisfactory result is somewhat reassuring. We now turn to the second part
of $\langle\zeta(t)\rangle_{\rm 1PI}$ illustrated in Figure 4(b). First, we
dispose of the diagram labelled (v).  In this diagram,
${\cal D}^{\zeta}_{\alpha\beta}$ acts on the internal propagator to produce
$\delta(0)$, and this contribution is precisely cancelled by the Jacobian
counterterm of diagram (iii).  In order to set
$\langle\zeta(t)\rangle_{\rm 1PI}$ equal to zero, we would like the remaining
contributions to sum to an expression of the form
$\int z_{\alpha}(t')g^{\zeta}_{\alpha\beta}(t',t)dt'$, so that $z_{\alpha}(t)$
can be set to zero.  The obstacle to this is diagram (vi), in which
${\cal D}^{\zeta}_{\alpha\beta}$ acts on the external propagator to produce
$\delta(t-t')$. Now, this problem could be solved through an integration by
parts to make ${\cal D}^{\zeta}_{\alpha\beta}$ act on the bubble to its left,
were it not for the boundary terms at $t'=t_0$, whose interpretation is a little
unclear.  It was to facilitate the handling of this integration by parts that we
introduced the counterterm proportional to $X_{\alpha}(t)$ in (\ref{Ltdc}),
which is a total derivative and contributes diagram (ii) of Figure 4(b).  In
effect, we can now perform the integration by parts without incurring boundary
terms by choosing
\begin{equation}
X_1(t)=X_2(t)=-\frac{\lambda^{1/2}}{2v(t)}I^{\zeta}_1(t)\,,
\end{equation}
where
\begin{equation}
I^{\zeta}_1(t)=\int\frac{d^3k}{(2\pi)^3}h^{\zeta}(t,t;k)
\label{i1def}
\end{equation}
is (with the spatial momentum now made explicit) the bubble contained in
diagrams (iv) and (vi). With this choice, the term
$X_{\alpha}{\cal D}^{\zeta}_{\alpha\beta}\zeta_{\beta}$ cancels diagram (iv)
while $-\zeta_{\alpha}{\cal D}^{\zeta}_{\alpha\beta}X_{\beta}$ supplies the
result of the integration by parts.  The counterterm was, of course, subtracted
off again in ${\cal L}_{\rm td}$ (equation (\ref{Ltd})), which we subsequently
assumed could be neglected. This assumption is now seen to amount to a
prescription for handling the present integration by parts and other similar
ones. In the present case, we take this prescription to be justified
{\it a posteriori} by the fact that we can now set $\langle\zeta(t)\rangle=0$
and hence obtain a sensible equation of motion for $v(t)$.  Including the
remaining diagram (i) (which represents the counterterm (\ref{Lct1})), the
equation of motion at this order is
\begin{eqnarray}
\left(\ddot{v}-m^2v+{1\over 6}v^3\right)&+&{3\over 2}\lambda
\left(\ddot{v}+M^2v\right)
\left(\frac{I^{\zeta}_1}{v^2}\right)\nonumber\\
&-&{1\over 2}\lambda\left(\partial^2_t+M^2\right)
\left(\frac{I^{\zeta}_1}{v}\right)=0\,.
\label{eqnofmotion}
\end{eqnarray}


\subsection{The 2-point function
$\langle\phi_{\alpha}^2(t)\phi_{\beta}^2(t')\rangle$}

It is now necessary to address several issues concerning the connected 2-point
function $\langle\phi_{\alpha}^2(t)\phi_{\beta}^2(t')\rangle_c$.  Clearly, if
$t<t_0$ and $t'>t_0$, this object is equal to
$2U(t')\langle\phi_{\alpha}^2(t)\zeta_{\beta}(t')\rangle$, while if $t$ and
$t'$ are both greater than $t_0$, it is equal to
$4U(t)U(t')\langle\zeta_{\alpha}(t)\zeta_{\beta}(t')\rangle$. Continuity of this
function at $t_0$ will supply the initial conditions needed to specify
$g^{\zeta}_{\alpha\beta}(t,t')$ completely, and we also require a suitable
definition of the renormalized effective mass $M(t)$.  We shall deal with these
issues at the lowest order of our approximation scheme, but first a technical
question must be addressed.

\subsubsection{Anchored contributions to
$\langle\zeta_{\alpha}(t)\zeta_{\beta}(t')\rangle$\label{anchor}}

The functional ${\cal K}$ which gives rise to our various propagators contains a
contribution ${\cal K}_{\delta}$, given in (\ref{Kdelta}) which is not included
in Figure 2.  In the toy calculation of section \ref{toy}, the analogous
quantity was found to cancel exactly another singular contribution (equation
(\ref{canceldelta})), provided that the propagators satisfied continuity
conditions which were expected on other grounds.  We now find a similar
cancellation involving the composite propagator of Figure 2(h), which is the
lowest-order anchored contribution to
$\langle\zeta_{\alpha}(t)\zeta_{\beta}(t')\rangle$. The internal lines in this
propagator correspond to the expression
\begin{equation}
\chi^{\phi}_{\alpha\beta}(t,t';k)=\int\frac{d^3k'}{(2\pi)^3}
g^{\phi}_{\alpha\beta}(t,t';k')g^{\phi}_{\alpha\beta}(t,t';k'+k)\,,
\end{equation}
and $2\chi^{\phi}_{\alpha\beta}(t,t';k)$ is precisely the lowest-order
contribution to $\langle\phi^2_{\alpha}(t)\phi^2_{\beta}(t')\rangle$.
Suppressing the momentum argument, the expression represented by Figure 2(h) is
\begin{equation}
\left.g^{\zeta}_{\alpha\gamma}(t,t_1)\tensor{d}_{\gamma\delta}(t_1)
\chi^{\phi}_{\delta\epsilon}(t_1,t_2)\tensor{d}_{\epsilon\sigma}(t_2)
g^{\zeta}_{\sigma\beta}(t_2,t')\right\vert_{t_1=t_2=t_0}\,.
\end{equation}
As indicated above, this expression is ill-defined, because it involves the
quantity $\left.\partial_{t_1}\partial_{t_2}\chi^{\phi}_{\alpha\beta}(t_1,t_2)
\right\vert_{t_1=t_2=t_0}$. However, we can use (\ref{gdgeqg}) (with
$\epsilon_{\alpha\beta}\tensor{\partial}_t$ replaced by
$\tensor{d}_{\alpha\beta}$) to find that this contribution to
${\cal K}_{\rm tr}$ is exactly cancelled by ${\cal K}_{\delta}$, provided that
the boundary conditions on $g^{\zeta}_{\alpha\beta}(t,t')$ satisfy
\begin{equation}
2\chi^{\phi}_{\alpha\beta}(t,t') =_0 4U(t)U(t')g^{\zeta}_{\alpha\beta}(t,t')\,.
\label{chidef}
\end{equation}
The notation $=_0$ here indicates that the quantities on the left and right,
together with their derivatives with respect to $t$ and $t'$ are equal at
$t=t'=t_0$.  If this prescription (which is consistent with the boundary
conditions to be discussed below) is adopted, then we may delete the propagator
of Figure 2(h).  This matter will, however, need to be reconsidered when
we come to discuss renormalization in section \ref{RENORMALIZATION}.

\subsubsection{Continuity of the 2-point function}

At the lowest order of approximation, the 2-point function 
$\langle\phi_{\alpha}^2(t)\phi_{\beta}^2(t')\rangle$ is represented by one of
the diagrams shown in Figure 5, depending on whether its time arguments are
greater or smaller than $t_0$. The anchored vertex in diagram 5(b) involves
$U(t_0)$ and $\dot{U}(t_0)$, which we expand in powers of $\lambda$, retaining
only the leading terms
\begin{eqnarray}
\left[U^{(0)}(t)\right]^2&=&
I^{\phi}_1(t)\equiv\int\frac{d^3k}{(2\pi)^3}h^{\phi}(t,t)\,,
\nonumber\\
U_0&\equiv&U^{(0)}(t_0)\,;\quad \dot{U}_0
=\left.\frac{d}{dt}U^{(0)}(t)\right\vert_{t=t_0}\,.
\label{U0def}
\end{eqnarray}
Because of the propagator structure exhibited in (\ref{galphabeta}), this
diagram can be expressed in the form
\begin{eqnarray}
g^{\phi\zeta}_{\alpha\beta}(&&t,t';k)
=-i\int\frac{d^3k'}{(2\pi)^3}\left[\frac{h^{\phi}_{\alpha}
(t_1,t;k')h^{\phi}_{\alpha}(t_1,t;k'+k)}{U^{(0)}(t_1)}\right]\nonumber\\
&&\times\left(\tensor{\partial}_{t_1}-\gamma^{\zeta}(t_1)\right)
\left.\left[h^{\zeta}(t',t_1;k)-{h^{\zeta}}^*(t',t_1;k)\right]
\right\vert_{t_1=t_0}.
\nonumber\\&&
\end{eqnarray}
In order to display the boundary conditions on the 2-point function succinctly,
we define
\begin{eqnarray*}
R_{\alpha}(t;k)&=&\int\frac{d^3k'}{(2\pi)^3}h^{\phi}_{\alpha}(t_0,t;k')
h^{\phi}_{\alpha}(t_0,t;k'+k)\,,\\
S_{\alpha}(t;k)&=&\partial_{t_1}\left.\int\frac{d^3k'}{(2\pi)^3}
h^{\phi}_{\alpha}(t_1,t;k')
h^{\phi}_{\alpha}(t_1,t;k'+k)\right\vert_{t_1=t_0},\\
P(t;k)&=&h^{\zeta}(t,t_0;k)-{h^{\zeta}}^*(t,t_0;k)\,,\\
Q(t;k)&=&\left(\partial_{t_1}-\gamma^{\zeta}_k(t_1)\right)\\
&&\quad\times
\left.\left[h^{\zeta}(t,t_1;k)-{h^{\zeta}}^*(t,t_1;k)\right]
\right\vert_{t_1=t_0},
\end{eqnarray*}
which satisfy, in particular,
\begin{eqnarray}
P(t_0)&=&0\,,\phantom{-i}\quad Q(t_0)=i\label{pq}\\
\dot{P}(t_0)&=&-i\,,\phantom{0}\quad \dot{Q}(t_0)=0\label{pqdot}
\end{eqnarray}
\begin{eqnarray}
R_0(k)&\equiv&R_1(t_0;k)=R_2(t_0;k)=R_0^*(k)\,,\label{R0def}\\
S_0(k)&\equiv&S_1(t_0;k)=S_2^*(t_0;k)\,.
\end{eqnarray}
In terms of these, we have
\begin{eqnarray}
g^{\phi\zeta}_{\alpha\beta}(t,t')=-\frac{i}{U_0}&&
\left[\vphantom{\frac{\dot{U}_0}{U_0}}\right.
R_{\alpha}(t)Q(t')\nonumber\\
&&\left.+\left(\frac{\dot{U}_0}{U_0}R_{\alpha}(t)-S_{\alpha}(t)\right)
P(t')\right]\,,
\end{eqnarray}
\begin{eqnarray}
\langle\phi^2_{\alpha}(t)\phi^2_{\beta}(t_0)\rangle&=&2R_{\alpha}(t)\,,\\
\left.\partial_{t'}\langle\phi^2_{\alpha}(t)\phi^2_{\beta}(t')\rangle
\right\vert_{t'=t_0}&=&2S_{\alpha}(t)\,.
\end{eqnarray}

In principle, the function $\langle\phi^2_{\alpha}(t)\phi^2_{\beta}(t')\rangle$
should be perfectly smooth as its arguments pass through $t_0$.  We will
actually attempt to impose the rather weaker conditions that the function itself
and its first derivatives are continuous at $t_0$.  At lowest order, these
conditions read
\begin{eqnarray}
g^{\phi\zeta}_{\alpha\beta}(t,t_0)&=&\frac{1}{2U_0}
\langle\phi^2_{\alpha}(t)\phi^2_{\beta}(t_0)\rangle\,,
\label{bc1}\\
g^{\phi\zeta}_{\alpha\beta}(t_0,t')&=&2U_0g^{\zeta}_{\alpha\beta}(t_0,t')\,,
\label{bc2}\\
\left.\partial_{t'}g^{\phi\zeta}_{\alpha\beta}(t,t')\right\vert_{t'=t_0}&=&
\partial_{t'}\left.\left[\langle\phi^2_{\alpha}(t)\phi^2_{\beta}(t')
\rangle\frac{1}{2U^{(0)}(t')}
\right]\right\vert_{t'=t_0}\,,\nonumber\\&&
\label{bc3}\\
\left.\partial_tg^{\phi\zeta}_{\alpha\beta}(t,t')\right\vert_{t=t_0}&=&
\left.\partial_t\left[2U^{(0)}(t)g^{\zeta}_{\alpha\beta}(t,t')\right]
\right\vert_{t=t_0}\,.
\label{bc4}
\end{eqnarray}
Of these, (\ref{bc1}) and (\ref{bc3}) are satisfied identically, because of
(\ref{pq}) and (\ref{pqdot}). The condition (\ref{bc2}) is satisfied provided
that
\begin{equation}
{\rm Im}\,S_0=-U_0^2
\label{ims}
\end{equation}
and
\begin{equation}
N^{\zeta}(t_0)=\frac{i}{U_0^2}\left[{\rm Re}\,S_0
- \left(Z-{\textstyle{1\over{2}}}
\gamma^{\zeta}(t_0)\right)R_0\right]\,,
\label{ninit}
\end{equation}
where
\begin{equation}
Z=\frac{\hbox{${\dot{f}}^{\zeta}$}^*(t_0)}{{f^{\zeta}}^*(t_0)}
+\frac{\dot{U}_0}{U_0}
\label{zinn}
\end{equation}
and $f^{\zeta}_k(t)$ is the mode function introduced in (\ref{modeequation})
that appears in $g^{\zeta}_{\alpha\beta}$.  Explicit calculation shows that
(\ref{ims}) is indeed true and (\ref{ninit}) of course provides one of the
initial values that we seek. The last condition (\ref{bc4}) is satisfied if
\begin{equation}
\dot{N}^{\zeta}(t_0)=\frac{i}{U_0^2}\left[{\rm Re}\,\dot{S}_0-2Z{\rm Re}\,S_0
+\left(Z^2-{\textstyle{1\over 4}}{\gamma^{\zeta}}^2(t_0)\right)R_0\right],
\label{ndotinit}
\end{equation}
with $\dot{S}_0=\dot{S}_1(t_0)=\dot{S}_2^*(t_0)$, and if also
\begin{equation}
{\rm Im}\,\dot{S}_0=-U_0^2\gamma^{\zeta}(t_0)\,.
\label{imsdot}
\end{equation}
The second of our initial values for the $\zeta$ propagator is, of course
provided by (\ref{ndotinit}), but the status of (\ref{imsdot}) is less clear.
Evaluating this equation explicitly, we obtain a moderately plausible relation
between the damping rates for $\phi$ and $\zeta$, namely
\begin{equation}
\int\frac{d^3k'}{(2\pi)^3}\gamma^{\phi}_{k'}(t_0)h^{\phi}(t_0,t_0;k'+k)
=U_0^2\gamma^{\zeta}_k(t_0)\,.
\label{gammas}
\end{equation}
However, this relation is not automatically satisfied.  Apparently, it implies
a constraint on the prescriptions used to define these damping rates, but we are
not able to say whether a prescription of the kind described in
\cite{lawrie1989} would naturally satisfy this constraint. In any case, when we
come to consider renormalization, we shall find that the boundary conditions
derived here require a significant modification.  We note, though, that in this
unrenormalized form, the initial values $N^{\zeta}(t_0)$ and
$\dot{N}^{\zeta}(t_0)$ do satisfy the condition (\ref{subsnequation}).

\subsubsection{The 2-point function
$\langle\zeta_{\alpha}(t)\zeta_{\beta}(t')\rangle$}

The last step required to specify the Feynman rules completely is to find a
suitable definition of the renormalized $\zeta$ mass $M(t)$.  To do this, we
calculate the 2-point function
$\langle\zeta_{\alpha}(t)\zeta_{\beta}(t')\rangle$ at first order in $\lambda$.
It is the sum of the eighteen diagrams shown in Figure 6 (recalling from our
earlier discussion that the composite propagator of Figure 2(h) can be ignored).
The computation of these diagrams is straightforward, except for Figure 6(s),
in which ${\cal D}_{\alpha\beta}$ acts twice on the same internal propagator.
In order to evaluate this diagram we have (i) ignored some contributions from
the dissipative coefficients $\alpha^{\zeta}$ and $\gamma^{\zeta}$ which can
formally be regarded as of higher order; (ii) adopted the following
prescription for the equal-time limit of derivatives of the propagator:
\begin{eqnarray*}
\left.\partial_tg^{\zeta}_{\alpha\alpha}(t,t')\right\vert_{t=t'}
&=&{\textstyle{1\over 2}}\left(\lim_{t'\uparrow t }
+\lim_{t'\downarrow t}\right)
\partial_tg^{\zeta}_{\alpha\alpha}(t,t')\\
&=&{\textstyle{1\over 2}}\left.\left[\partial_th^{\zeta}(t,t')
+\partial_{t'}h^{\zeta}(t,t')
\right]\right\vert_{t=t'};
\end{eqnarray*}
and (iii) identified the second counterterm in (\ref{Ltdc}) as
\begin{equation}
Y_1(t)=Y_2(t)=\frac{\lambda}{8}\frac{I_1^{\zeta}(t)}{v^2(t)}\,,
\end{equation}
where $I_1^{\zeta}(t)$ was defined in (\ref{i1def}), so as to facilitate an
integration by parts. The result we obtain is consistent, at this order of
approximation, with the structure displayed in (\ref{zetaandpsi}). Specifically,
we write it as
\begin{eqnarray}
G^{\zeta}_{\alpha\beta}(t,t')&=&\frac{\sigma(t)}{v(t)}
G^{(\psi\psi)}_{\alpha\beta}(t,t')
\frac{\sigma(t')}{v(t')}\nonumber\\
&&+\frac{\lambda^{1/2}}{2}\frac{\sigma(t)}{v(t)}
G^{(\psi\psi^2)}_{\alpha\beta}(t,t')
\frac{1}{v(t')}\nonumber\\
&&+\frac{\lambda^{1/2}}{2}\frac{1}{v(t)}G^{(\psi^2\psi)}_{\alpha\beta}(t,t')
\frac{\sigma(t')}{v(t')}\nonumber\\
&&+\frac{\lambda}{4}\frac{1}{v(t)}G^{(\psi^2\psi^2)}_{\alpha\beta}(t,t')
\frac{1}{v(t')}\,,
\label{zetawithpsi}
\end{eqnarray}
where the quantity
\begin{equation}
\sigma(t)=v(t)\left[1-\frac{\lambda}{2}\frac{I^{\zeta}_1(t)}{v^2(t)}
+O(\lambda^2)\right]
\label{sigmadef}
\end{equation}
might loosely be interpreted as corresponding to the peaks of a probability
density of the kind sketched in Figure 1(d). In the first term of
(\ref{zetawithpsi}), with which we are principally concerned, we have
\begin{eqnarray}
&&G^{(\psi\psi)}_{\alpha\beta}(t,t')\nonumber\\
&&\quad =g^{\zeta}_{\alpha\beta}(t,t')
+i\int_{t_0}^{t_f}dt''g^{\zeta}_{\alpha\gamma}(t,t'')
\epsilon_{\gamma\delta}{\cal A}(t'')
g^{\zeta}_{\delta\beta}(t'',t')\nonumber\\
&&\qquad+\int_{t_0}^{t_f}dt''dt'''g^{\zeta}_{\alpha\gamma}(t,t'')
\epsilon_{\gamma\delta}
{\cal B}_{\delta\epsilon}(t'',t''')\epsilon_{\epsilon\lambda}
g^{\zeta}_{\lambda\beta}(t''',t')\,,\nonumber\\
&&\label{Gpsipsi}
\end{eqnarray}
with auxiliary functions ${\cal A}(t)$ and ${\cal B}_{\alpha\beta}(t,t')$ which
will be specified shortly. It will be seen, for example, that the term
$(\sigma/v)_tg^{\zeta}_{\alpha\beta}(t,t')(\sigma/v)_{t'}$ arises from the sum
of diagrams (a), (g) and (h) of Figure 6, together with a 2-loop diagram which
is easily seen to be present at order $\lambda^2$.  As always in the
closed-time-path formalism, the signs $\epsilon_{\alpha\beta}$ attached to the
vertices and the structure (\ref{galphabeta}) of the propagators ensure
causality, in the sense that the integrands in (\ref{Gpsipsi}) vanish whenever
$t''$ or $t'''$ is greater than both $t$ and $t'$. The auxiliary functions are
given by
\begin{eqnarray}
{\cal A}(t)\,&&=\left(\frac{\ddot{v}+M^2v-{\textstyle{1\over 3}}v^3}
{v}\right)_t
+\frac{3\lambda}{2}\left(\frac{\ddot{v}+M^2v}{v^3}\right)_tI^{\zeta}_1(t)
\nonumber\\
&&-\frac{\lambda}{2v(t)}\left(\partial_t^2+M^2\right)
\left(\frac{I^{\zeta}_1}{v}\right)_t
+\frac{9\lambda}{2}\left(\frac{\ddot{v}+M^2v}{v^2}\right)^2_t
\bar{I}_2\,,\nonumber\\
&&
\label{Aoft}
\end{eqnarray}
and
\begin{eqnarray}
{\cal B}_{\alpha\beta}(t,t')&&\,
=-\frac{9\lambda}{2}\left(\frac{\ddot{v}+M^2v}{v^2}\right)_t
\nonumber\\
&&\times\left[I^{\zeta}_2(t,t')_{\alpha\beta}
+i\epsilon_{\alpha\beta}\delta(t-t')\bar{I}_2
\right]\left(\frac{\ddot{v}+M^2v}{v^2}\right)_{t'},\nonumber\\&&
\label{Balphabeta}
\end{eqnarray}
with
\begin{equation}
I_2^{\zeta}(t,t';k)_{\alpha\beta}=\int\frac{d^3k'}{(2\pi)^3}
g^{\zeta}_{\alpha\beta}(t,t';k')
g^{\zeta}_{\alpha\beta}(t,t';k'+k)\,.
\end{equation}
The terms proportional to $\bar{I}_2$ cancel out in (\ref{Gpsipsi}) and have
been inserted for the purpose of renormalization, which is considered below.
For completeness, we record that the remaining functions are
\begin{eqnarray*}
G^{(\psi\psi^2)}_{\alpha\beta}(t,t')&=&-3i\lambda^{1/2}\int_{t_0}^{t_f}dt''
g^{\zeta}_{\alpha\gamma}(t,t'')\epsilon_{\gamma\delta}\\
&&\quad\times\left(\frac{\ddot{v}+M^2v}{v^2}\right)_{t''}
I^{\zeta}_2(t'',t')_{\delta\beta}\\
G^{(\psi^2\psi)}_{\alpha\beta}(t,t')&=&
-3i\lambda^{1/2}\int_{t_0}^{t_f}dt''
I^{\zeta}_2(t,t'')_{\alpha\gamma}\\
&&\quad\times\left(\frac{\ddot{v}+M^2v}{v^2}\right)_{t''}
\epsilon_{\gamma\delta}g^{\zeta}_{\delta\beta}(t'',t')\\
G^{(\psi^2\psi^2)}_{\alpha\beta}(t,t')&=&2I^{\zeta}_2(t,t')_{\alpha\beta}\,.
\end{eqnarray*}
These are easily seen to correspond at lowest order to the correlators indicated
in (\ref{zetaandpsi}).

Our renormalized mass $M(t)$ will now be defined by requiring ${\cal A}(t)=0$.
In the nonequilibrium theory, there is no clear analogue of the pole of a
zero-temperature propagator in Minkowski spacetime, which unambiguously
identifies the mass of a stable particle.  Here, our rationale is, rather, to
optimise $g^{\zeta}_{\alpha\beta}$ as an approximation to the full 2-point
function by using the mass counterterm (\ref{massct}) to cancel as much as
possible of the higher order corrections.  Although the function ${\cal A}(t)$
was identified by eye from our explicit $O(\lambda)$ result, we assume that
this condition actually makes sense to all orders, so the definition of $M(t)$
is an implicit one, which can be realised in practice only to a given order of
approximation.  Setting the expression (\ref{Aoft}) to zero appears to give a
second equation of motion for $v(t)$ which is similar, but not identical, to
that already found in (\ref{eqnofmotion}) by requiring
$\langle\zeta(t)\rangle=0$.  In fact, these two
equations together yield both the equation of motion for $v(t)$ and the
relation between $M(t)$ and the bare mass $m(t)$.  Consistently ignoring
terms of order $\lambda^2$, we obtain the equation of motion
\begin{eqnarray}
\left(\partial_t^2+M^2\right)&&\left[v-\frac{\lambda}{2}
\frac{I_1^{\zeta}}{v}
+O(\lambda^2)\right]\nonumber\\
&&=\frac{v^3}{3}\left[1-\frac{3\lambda}{2}\frac{I_1^{\zeta}}{v^2}
-\frac{3}{2}\lambda\bar{I}_2
+O(\lambda^2)\right]
\label{eqofm}
\end{eqnarray}
and the gap equation
\begin{equation}
m^2(t)=-M^2(t)+{\textstyle{1\over 2}}v^2(t)\left[1-\lambda \bar{I}_2\right]
+O(\lambda^2)\,.
\label{massrelation}
\end{equation}
This gap equation expresses the bare mass $m(t)$, which occurs only in the
counterterm (\ref{linearct}), as a function of $M(t)$, $v(t)$ and $\lambda$.
For a given $m(t)$, these are two equations to be solved simultaneously for
$v(t)$ and $M(t)$.  In the specific application to an expanding universe, we
have $m^2(t)=a^2(t)m_0^2$, and one must simultaneously solve Einstein's field
equations for the scale factor $a(t)$.


\section{Renormalized equations of motion and initial conditions
\label{renormalization}}

The equations of motion we have just derived involve integrals $I^{\zeta}_1(t)$
and $I^{\zeta}_2(t,t';k)_{\alpha\beta}$ which are ultraviolet divergent, and
many more divergent integrals are of course to be expected at higher orders.
Of necessity, the nonequilibrium theory is formulated in terms of quantities
which are defined implicitly as the solutions of differential equations or of
self-consistency relations.  This system of equations, while ultimately
susceptible of numerical solution when truncated at some (no doubt low) order of
perturbation theory, is far too complicated to admit of any exact analytical
solution, and a proof that they can be renormalized so as to remove ultraviolet
divergences at all orders is currently beyond the ingenuity of this author.  A
numerical investigation at the order of approximation we have considered
explicitly ought to be feasible, provided that a suitable renormalization
prescription can be given.  Even at this level, we have in hand only formal
expressions for the propagators $g_{ab}(t,t')$ in terms of mode functions
$f_k(t)$ and generalized ``occupation numbers'' represented by the functions
$N_k(t)$, which must be found from numerical solution of (\ref{modeequation})
and (\ref{nequation}), so an unambiguous proof of renormalizability is rather
difficult.  Here, we suggest a renormalization prescription which might be
applied in the context of a numerical computation involving a spatial momentum
cutoff $\Lambda$, which reflects the renormalizations that are well known to
work in the Minkowski-space theory.  Since ultraviolet divergences are a feature
of the short-distance and short-time properties of the theory, one expects that
they should be essentially independent of the nonequilibrium state, and that
these renormalizations should be adequate.  In practice, one hopes that
numerical computations would approach finite limits when $\Lambda\to\infty$, and
we shall offer circumstantial grounds for optimism that this should indeed be so.

\subsection{Large-$k$ behaviour of propagators and 1-loop integrals}

We need to estimate the large-$k$ behaviour of our propagators, and do so using
a method similar to that described, for example, in \cite{boyanovsky1994}.
The usual propagators of the equilibrium theory are obtained from our
$g_{\alpha\beta}(t,t')$, by setting $N_k(t)=2n_k+1$, where $n_k$ is the
Bose-Einstein distribution, which falls off exponentially at large $k$.  When
time evolution is sufficiently slow, the dissipative coefficients $\alpha_k(t)$
and $\gamma_k(t)$ are given by scattering integrals \cite{lawrie1989}, which are
convolutions of the $n_k$, and also exponentially small at large $k$.  For
present purposes therefore, we assume that it is sufficient to set $N_k(t)=1$
and $\alpha_k(t)=\gamma_k(t)=0$.  Then, from (\ref{littleh}), we need only to
estimate $h(t,t';k)\simeq{1\over 2}f_k(t)f^*_k(t')$, with mode functions obeying
\begin{equation}
\left[\partial_t^2+k^2+M^2(t)\right]f_k(t)=0
\label{newmode}
\end{equation}
and the Wronskian condition (\ref{wronskian}).  The general solution may be
written as
\begin{equation}
f_k(t)=\frac{e^{i\theta(k)}}{\sqrt{2\Omega_k(t)}}\exp\left[
-i\int_0^t\Omega_k(t')dt'\right]\,,
\end{equation}
where $\theta(k)$ is an arbitrary, but time-independent, function of $k$ which
does not affect the propagator and will therefore be ignored.  We will choose
(as is always possible) a time-dependent frequency which behaves as
$\Omega_k(t)=k+O(k^0)$ for large $k$.  In this case, we obtain the large-$k$
expansion
\begin{eqnarray}
f_k(t)&&f_k^*(t')=\frac{e^{-ik(t-t')}}{2k}\nonumber\\
&&\times\left[1-\frac{i}{2k}W(t,t')-\frac{1}{4k^2}X(t,t')+O(k^{-3})\right],
\end{eqnarray}
where
\begin{equation}
W(t,t')=\int_{t'}^tM^2(t'')dt''
\end{equation}
and $X(t,t')=M^2(t)+M^2(t')+\frac{1}{2}W^2(t,t')$.  This expansion can be used
to isolate the ultraviolet divergences of the two integrals $I_1^{\zeta}(t)$ and
$I_2^{\zeta}(t,t';k)$.  We impose a cutoff $\Lambda$ on {\it physical} momenta
$k_{\rm ph}=\vert\bbox{k}_{\rm ph}\vert<\Lambda$.  Up to this point, our
analysis has been entirely in terms of comoving coordinates and momenta and,
in principle, we should translate all of the above results into physical
coordinates in order to impose this cutoff systematically.  However, both
$I_1^{\zeta}(t)$ and the divergent part of $I_2^{\zeta}(t,t';k)$ are local in
time and, for our present purposes, it is sufficient simply to convert the
physical cutoff to a comoving cutoff $k<a(t)\Lambda$. Specifically, we find
\begin{equation}
I_1^{\zeta}(t)=\frac{1}{8\pi^2}\left[\left(a(t)\Lambda\vphantom{^2}\right)^2
-M^2(t)\ln\left(a(t)\Lambda\vphantom{^2}\right)\right]+\cdots\ ,
\end{equation}
\begin{eqnarray}
I_2^{\zeta}(t,t';k)_{\alpha\beta}
&=&-i\delta_{\alpha\beta}\epsilon_{\alpha}\delta(t-t')\bar{I}_2(t)+\cdots\ ,\\
\bar{I}_2(t)&=&\frac{1}{8\pi^2}\ln\left(a(t)\Lambda\vphantom{^2}\right)\,,
\label{I2bar}
\end{eqnarray}
where the ellipsis represents ultraviolet-finite contributions.  The quantity
$\bar{I}_2(t)$ here is that appearing in (\ref{Aoft}) and (\ref{Balphabeta})
and we see, in particular, that the net integral in (\ref{Balphabeta}) is
finite.

\subsection{Renormalized equation of motion and gap equation}

We hope, of course, that both the equation of motion (\ref{eqofm}) and the gap
equation (\ref{massrelation}) can be expressed in a form which is free of
ultraviolet divergences. To this end, we introduce the physical particle mass
$\hat{m}$, which locates the pole of the Minkowski-space propagator in the
broken-symmetry vacuum.  It is related to the bare mass $m_0$ by
\begin{eqnarray}
m_0^2=\frac{\lambda}{16\pi^2}\Lambda^2+\frac{1}{2}\hat{m}^2
\left\{\vphantom{\frac{\Lambda}{6}}
\right.\!\!1+\frac{\lambda}{16\pi^2}
\left[\vphantom{\frac{\Lambda}{6}}\right.
\ln&&\left(\frac{\Lambda}{\hat{m}}\right)+c\left.\left.
\vphantom{\frac{\Lambda}{6}}\right]\right\}\nonumber\\
&&\quad +\,O(\lambda^2)\,,
\end{eqnarray}
with $c=1+\ln 2-\sqrt{3}\pi/2$. (The 1-loop integrals used to obtain this
relation are, of course, Minkowski-space integrals involving a physical
3-momentum $\bbox{k}_{\rm ph}$, whose magnitude is cut off at the value
$\Lambda$.)  In the Minkowski-space theory, Green's functions involving the
operator $\phi^2$ are rendered finite by a combination of additive and
multiplicative renormalizations. Guided by the standard theory of these
renormalizations, we surmise that a renormalized expectation value
$v_{\rm R}^2(t)=\langle\phi^2\rangle_{\rm R}$ can be defined, at the order of
approximation we are using, by
\begin{equation}
\frac{v^2}{\lambda}=\frac{(a\Lambda)^2}{8\pi^2}+\frac{1}{2}\hat{m}^2a^2
\left(1+\frac{\lambda_R}{16\pi^2}\bar{c}\right)\bar{I}_2+Z_{\phi^2}
\left(\frac{v_{\rm R}^2}
{\lambda_{\rm R}}\right),
\end{equation}
where $Z_{\phi^2}=1-{1\over 2}\lambda_{\rm R}\bar{I}_2+O(\lambda_{\rm R}^2)$
is the multiplicative renormalization factor and
$\bar{c}(t)=c-\ln(a(t)\hat{m})$.  The renormalized coupling constant
$\lambda_{\rm R}$ is defined by $\lambda=Z_{\lambda}\lambda_{\rm R}$, with
$Z_{\lambda}=1+{3\over 2}\lambda_{\rm R}\bar{I}_2+O(\lambda_{\rm R}^2)$.  With
these definitions, we indeed obtain a renormalized equation of motion
\begin{eqnarray}
\left(\partial_t^2+M^2\right)&&\left[v_{\rm R}-\frac{\lambda_{\rm R}}{2}
\frac{\tilde{I}_1^{\zeta}}{v_{\rm R}}+O(\lambda_{\rm R}^2)\right]\nonumber\\
&&=\frac{v_{\rm R}^3}{3}\left[1-\frac{3\lambda_{\rm R}}{2}
\frac{\tilde{I}_1^{\zeta}}{v_{\rm R}^2}
+O(\lambda_{\rm R}^2)\right]
\label{reneqofm}
\end{eqnarray}
and a renormalized gap equation
\begin{equation}
M^2(t)=\frac{1}{2}v_{\rm R}^2-\frac{1}{2}\hat{m}^2a^2(t)\left(1
+\frac{\lambda_{\rm R}}{16\pi^2}
\bar{c}\right)+O(\lambda_{\rm R}^2)\,,
\label{rengap}
\end{equation}
where
\begin{equation}
\tilde{I}_1^{\zeta}(t)=I_1^{\zeta}(t)-\left[\left(a(t)
\Lambda\vphantom{^2}\right)^2
-M^2(t)\ln\left(a(t)\Lambda\vphantom{^2}\right)\right]/(8\pi^2)\,.
\label{Itilde}
\end{equation}
If our assumptions about the large-$k$ behaviour of the integrals are correct,
then these are finite relations between renormalized quantities. (To simplify
matters, we have neglected a term in (\ref{reneqofm}) involving the time
derivatives of $\bar{I}_2$ which is finite, but probably an artifact of our
regularization procedure.) 

For the symmetric state which exists prior to $t_0$, the renormalized mass
$\mu(t)$ is defined in a similar manner to $M(t)$ and satifies the gap equation 
\begin{equation}
\mu^2(t)=\frac{\lambda_{\rm R}}{2}\tilde{I}_1^{\phi}(t)
-\frac{1}{2}a^2(t)\hat{m}^2\left(1+\frac{\lambda_{\rm R}}{16\pi^2}
\bar{c}\right)+O(\lambda_{\rm R}^2)\,,
\label{mugap}
\end{equation}
where $\tilde{I}^{\phi}_1(t)$ is defined in the same way as
$\tilde{I}^{\zeta}_1(t)$, but using the $\phi$ propagator and mass $\mu(t)$.

\subsection{Initial conditions at $t=t_0$\label{init}}

The equation of motion (\ref{reneqofm}) needs, of course, the initial values
$v_{\rm R}(t_0)$ and $\dot{v}_{\rm R}(t_0)$ and these are obtained from the
continuity of $\langle\phi^2(t)\rangle$ and  $d\langle\phi^2(t)\rangle/dt$ at
$t=t_0$.  This expectation value is given by
$\langle\phi^2(t)\rangle =I_1^{\phi}+O(\lambda)$ just before $t_0$ and by
definition it is equal to $\lambda^{-1}v^2(t)$ just after $t_0$.  In terms of
$v_{\rm R}$, the continuity conditions are (in the notation of (\ref{chidef}))
\begin{equation}
v_{\rm R}^2(t)=_0 \lambda_{\rm R}\tilde{I}_1^{\phi}(t)+O(\lambda_{\rm R}^2)\,.
\label{vrinit}
\end{equation}
As a consequence, we find that $M(t_0)=\mu(t_0)$ at this order, suggesting the
satisfactory possibility that the excitations of the state at $t_0$ described by
$\phi$ and by $\zeta$ are essentially the same.

Our final task is to investigate how the initial conditions (\ref{ninit}) and
(\ref{ndotinit}) for the generalized occupation numbers $N_k^{\zeta}(t)$ and
$\dot{N}_k^{\zeta}(t)$ might be renormalized. It turns out that a far-reaching
modification of the strategy used to obtain these conditions is needed.  Indeed,
(\ref{ninit}) has a potentially disastrous conseqence, since it implies that
\begin{eqnarray}
g_{11}^{\zeta}(t_0,t_0;k)&=&\frac{1}{2}f^{\zeta}_k(t_0){f^{\zeta}_k}^*(t_0)
\left[N_k^{\zeta}(t_0)+{N_k^{\zeta}}^*(t_0)\right]\nonumber\\
&=&\frac{R_0(k)}{2U_0^2}\,.
\end{eqnarray}
On integrating this relation we discover that
\begin{eqnarray}
I_1^{\zeta}(t_0)&=&\int\frac{d^3k}{(2\pi)^3}g^{\zeta}_{11}(t_0,t_0;k)=
\frac{1}{2U_0^2}\int\frac{d^3k}{(2\pi)^3}R_0(k)\nonumber\\
&=&\frac{1}{2}I_1^{\phi}(t_0)\,.
\label{badresult}
\end{eqnarray}
The renormalization programme we have outlined so far clearly requires that the
ultraviolet divergences of $I_1^{\zeta}$ and $I_1^{\phi}$ should be the same,
which is incompatible with (\ref{badresult}).  The problem arises from the fact
that we need to match two different perturbation series at $t_0$, whereas a
consistent renormalization scheme can be implemented only within a single
systematic expansion.  In view of this, we replace
$g^{\zeta}_{\alpha\beta}(t,t')$ in (\ref{bc2}) and (\ref{bc4}) with a function
$\bar{g}^{\zeta}_{\alpha\beta}(t,t')$ obtained by summing those contributions
to the 2-point function for $\zeta$ which are of order $\lambda^0$, when $v$ is
regarded as of order $\lambda^{1/2}$, as it is in the $\phi$-theory.  This
function is
\begin{equation}
\bar{g}^{\zeta}_{\alpha\beta}(t,t')=\frac{\bar{\sigma}(t)}
{v(t)}g^{\zeta}_{\alpha\beta}(t,t')
\frac{\bar{\sigma}(t')}{v(t')}+\frac{\lambda}{2}\frac{1}{v(t)}
I_2^{\zeta}(t,t')\frac{1}{v(t)}\,,
\end{equation}
where
\begin{equation}
\bar{\sigma}(t)=\left[v^2(t)-\lambda I_1^{\zeta}(t)\right]^{1/2}
\label{sigmabardef}
\end{equation}
and $v(t)$ is understood to be evaluated at lowest order. On evaluating the
modified version of (\ref{bc2}) at $t'=t_0$, we find
\begin{equation}
\bar{\sigma}^2(t_0)g^{\zeta}_{11}(t_0,t_0;k)=\frac{\lambda}{2}
\left[I_2^{\phi}(t_0,t_0;k)
-I_2^{\zeta}(t_0,t_0;k)\right]\,,
\label{newncondition}
\end{equation}
and we want to find initial conditions for $g^{\zeta}$ for which this condition
holds. It clearly will hold if $\bar{\sigma}(t_0)=0$ and
$I_2^{\phi}(t_0,t_0;k)=I_2^{\zeta}(t_0,t_0;k)$, and this is in fact the only
possibility.  On integrating (\ref{newncondition}) and using in
(\ref{sigmabardef}) the fact that $v^2(t_0)=\lambda I_1^{\phi}(t_0)$, we obtain
\begin{equation}
\left[I_1^{\phi}(t_0)-I_1^{\zeta}(t_0)\right]I_1^{\zeta}(t_0)=\frac{1}{2}
\left[I_1^{\phi}(t_0)^2-I_1^{\zeta}(t_0)^2\right]\,,
\end{equation}
which implies that $I_1^{\phi}(t_0)=I_1^{\zeta}(t_0)$ and hence
$\bar{\sigma}(t_0)=0$.  Now, we found above that the $\phi$ and $\zeta$ masses
$\mu(t)$ and $M(t)$ are equal (at lowest order) at $t=t_0$, so we can choose the
same mode functions $f_k(t)$ for both propagators near $t=t_0$. Evidently,
(\ref{newncondition}) is satisfied if we also choose
$N_k^{\zeta}(t_0)=N_k^{\phi}(t_0)$ and, in fact, the modified versions of
(\ref{bc1}) - (\ref{bc4}) all hold at $t=t'=t_0$ if we choose
$\dot{N}_k^{\zeta}(t_0)=\dot{N}_k^{\phi}(t_0)$. We thus reach the apparently
satisfactory conclusion that the propagators $g^{\zeta}_{\alpha\beta}(t,t')$ and
$g^{\phi}_{\alpha\beta}(t,t')$ and their first derivatives are the same at
$t_0$, so that the $\phi$ and $\zeta$ excitations are essentially the same.
The requirement (\ref{subsnequation}) is of course satisfied for
$N_k^{\zeta}(t)$, since it is already satisfied for $N_k^{\phi}(t)$.  The new
boundary conditions do, of course, spoil the exact cancellation between the
composite propagator of figure 2(h) and the unwanted ${\cal K}_{\delta}$
discussed in section \ref{anchor}.  This propagator can be calculated explicitly
using the representation (\ref{galphabeta}) for the internal lines and setting
the anchored time arguments equal to $t_0$ after the required derivatives have
been performed. We find that derivatives of $\theta(t-t')$ produce terms
proportional to $\delta(0)$ which are again exactly cancelled by
${\cal K}_{\delta}$, but there are now residual terms.  These
contribute to the function ${\cal B}_{\alpha\beta}$ in (\ref{Gpsipsi}) and do
not enter into our explicit calculations.

At this point, then, we have a fully renormalized equation of motion and gap
equation and finite initial conditions for both the equation of motion and the
propagator $g^{\zeta}_{\alpha\beta}$.  To the first non-trivial order of our
approximation scheme, all the ingredients are to hand for the calculation of the
expectation values we want, which must necessarily be completed by numerical
methods.  This apparently satisfactory state of affairs conceals, however, an
uncomfortable feature of perturbation theory which must now be exposed.
If we define
\begin{equation}
\sigma_{\rm R}^2(t)=v_{\rm R}^2(t)-\lambda_{\rm R}\tilde{I}_1^{\zeta}(t)
+O(\lambda_{\rm R}^2)\,,
\label{sigmardef}
\end{equation}
then (\ref{reneqofm}) and (\ref{rengap}) are consistent at this order with
the pair of equations
\begin{equation}
\ddot{\sigma}_{\rm R} +M^2(t)\sigma_{\rm R}=\textstyle{\frac{1}{3}}
\sigma_{\rm R}^3+O(\lambda_{\rm R}^2)\,,
\label{sigmaeom}
\end{equation}
\begin{eqnarray}
M^2(t)=&&\textstyle{1\over 2}\sigma_{\rm R}^2(t)+\textstyle{1\over 2}
\lambda_{\rm R}\tilde{I}_1^{\zeta}(t)\nonumber\\
&&\quad-\textstyle{1\over 2}\hat{m}^2a^2(t)
\left[1+\textstyle{1\over 2}\lambda_{\rm R}\bar{c}(t)\right]
+O(\lambda_{\rm R}^2)\,.
\label{sigmagap}
\end{eqnarray}
Suppose, for the sake of argument, that $a(t)$ approaches the constant value
$a=1$ at late times. There are two steady-state solutions to (\ref{sigmaeom})
and (\ref{sigmagap}) to which the final state might correspond.  The first is
$\sigma_{\rm R}=\sqrt{3}M+O(\lambda_{\rm R}^2)$ and
$M^2=\hat{m}^2-{1\over 2}\lambda_{\rm R}\tilde{I}_1^{\zeta}
+O(\lambda_{\rm R}^2)$. This corresponds as in section \ref{sus} to a
superposition of states in which $\langle\phi\rangle=\pm\sigma_{\rm R}$, and
is the state that we might hope to see emerging. The second solution is
$\sigma_{\rm R}=0$ and $M^2=-{1\over 2}\hat{m}^2
+{1\over 2}\lambda_{\rm R}\tilde{I}_1^{\zeta}+O(\lambda_{\rm R}^2)$.  In terms
of the conventional description of spontaneous symmetry breaking in Minkowski
spacetime, this of course corresponds to the ill-defined perturbation theory
with $\langle\phi\rangle=0$.  Which of these situations will emerge from our
dynamical description depends crucially on how we treat our perturbative
expansions.  Using (\ref{sigmardef}) as it stands, the initial conditions we
have derived would imply that $\sigma_{\rm R}=0$ for all times ($\sigma_{\rm R}$
being essentially a renormalized version of the $\bar{\sigma}$ introduced in
(\ref{sigmabardef})), and hence that $v_{\rm R}^2=\lambda_{\rm R}\tilde{I}_1^{\zeta}
=\lambda_{\rm R}\tilde{I}_1^{\phi}$ at all times. Although we have only
first-order results in hand, this strongly suggests that that the $\zeta$-theory
would then simply reproduce the $\phi$-theory, but by a more complicated
route. This perturbation theory is unsatisfactory when $v_{\rm R}$ becomes
of order 1, as it eventually must do.  On the other hand, if we use the actual
$O(\lambda_{\rm R})$ result in (\ref{reneqofm}) to define
\begin{equation}
\sigma_{\rm R}=v_{\rm R}-\frac{\lambda_{\rm R}}{2}
\frac{\tilde{I}_1^{\zeta}}{v_{\rm R}}+O(\lambda_{\rm R}^2)
\label{newsigmardef}
\end{equation}
and truncate the expansion at this order, then $\sigma_{\rm R}$ and
$\dot{\sigma}_{\rm R}$ receive nonzero initial values at $t_0$ and should
evolve qualitatively in the expected manner. This ambiguity is inherent in
any perturbative approach to our problem, which necessarily requires us to match
the two different expansions about the states indicated schematically in
figures 1(c) and 1(d).  Intuitively, we would like to interpret the definition
(\ref{sigmardef}) as yielding $\sigma_{\rm R}=\langle\phi\rangle\equiv 0$ and
(\ref{newsigmardef}) as approximating the locations of the most probable values
of $\phi$ in figure 1(d).  In practical terms, the most appropriate strategy
seems to be that set out earlier.  Namely, we make no reference to $\sigma_R$,
but deal exclusively with $v_{\rm R}$, which can be calculated both in the
$\phi$-theory and in the $\zeta$-theory.  Within the $\zeta$-theory, we adhere
to a strict series expansion of (\ref{reneqofm}) and (\ref{rengap}) in powers
of $\lambda_{\rm R}$, treating the initial values obtained from the
$\phi$-theory as purely numerical input to these equations, even though the
numbers are obtained within the $\phi$-theory from an expansion in powers of
$\lambda_{\rm R}$.  In principle, at least, this yields a sequence of
approximations which can be pursued systematically to arbitrarily high orders.

\section{Discussion\label{discussion}}

In this paper, we have proposed a perturbative approximation scheme through
which quantum-field-theoretic expectation values might be estimated during the
course of a symmetry-breaking phase transition.  The scheme has two distinctive
features, both of which seem to be essential to this problem.  First, in order
that low-order calculations adequately reflect the evolving nonequilibrium
state, this state is described in terms of its own quasiparticle excitations.
Lowest-order propagators describe the propagation of these excitations and
incorporate approximations to their mass (or, more generally, their dispersion
relation) and damping rate, which can be obtained as the solutions of
appropriate self-consistency conditions. Second, because we deal with an exact
symmetry $\phi\leftrightarrow -\phi$, which is a symmetry both of the
Hamiltonian that governs time evolution and of the initial state, the scalar
field $\phi$ can never acquire a nonzero expectation value.  In fact, the
symmetry can never, properly speaking, be broken at all. In Minkowski spacetime,
nevertheless, the effects conventionally associated with a nonzero value
of $\langle\phi\rangle$ can be recovered in perturbation theory by dealing
instead with $\langle\phi^2\rangle$ and we have here generalised this
description to accommodate the state emerging dynamically from the phase
transition. (As noted in section \ref{INTRO}, the value of
$\langle\phi\rangle$ does not in itself provide a full characterisation of the
nonequilibrium state.  Nor, of course, does the value of $\langle\phi^2\rangle$,
but this quantity plays an indispensable role in the perturbative description
of this state.)

Inherent in any attempt to apply perturbation theory to this problem is the
difficulty that $\langle\phi^2\rangle$ is formally of order $\lambda^0$ in the
early, symmetrical state, but of order $\lambda^{-1}$ in the state that emerges
from the transition. This necessarily means that two inequivalent perturbation
series must be used, before and after a time that we have called $t_0$.
Intuitively, we suppose that before $t_0$, the most probable value of $\phi$ is
zero, while after $t_0$ the most probable values are, say,
$\pm\sigma/\sqrt{\lambda}$.  However, perturbation theory gives no clear
indication of the instant at which this bifurcation occurs, so in practice an
{\it ad hoc} choice of $t_0$, perhaps optimising in some way the matching of the
two approximations, would be required.  More importantly, perhaps, the above
difficulty leads to two specific shortcomings of our analysis which deserve some
emphasis.  One is that we had to perform in section \ref{pathintegral} a
somewhat singular transformation of the path integral from which expectation
values are computed, which led to ambiguities concerning the treatment of total
time derivatives and the meaning of derivatives of propagators when their time
arguments coincide at $t_0$.  We succeeded in resolving these ambiguities
sufficiently for the immediate purposes of the explicit calculations described
in sections \ref{EOMCON} and \ref{RENORMALIZATION}, but have not obtained any
well-founded general resolution.  It is possible that a sufficiently careful
derivation might avoid these ambiguities, but we have not yet been able to
achieve this.  The second shortcoming is that discussed in section \ref{init},
namely that our perturbative estimate of $v_{\rm R}(t)$ from the equation of
motion (\ref{reneqofm}) and the initial contitions (\ref{vrinit}) depend
crucially on how the perturbation series is organised.  We gave a prescription
which should apparently produce the expected type of behaviour, and can be
pursued systematically, but it is not unique.  We do not think that any
significantly better strategy is available within perturbation theory.
Possibly, some non-perturbative means of improving the matching between the
two expansions might be found, but we have not yet been able to do this.

Quite extensive numerical calculations are needed to extract concrete results
from the perturbation theory we have constructed, and we hope to report such
results in future publications.  In the absence of further approximations beyond
the perturbation expansion, the numerical problem involves the simultaneous
solution of several coupled nonlinear equations: the equation of motion
(\ref{reneqofm}); the gap equation (\ref{rengap}); the equations
(\ref{modeequation}) and (\ref{nequation}) for mode functions and generalized
occupation numbers, and two equations we have not given explicitly for the
dissipative coefficients $\alpha_k(t)$ and $\gamma_k(t)$, which involve at least
2-loop integrals, as detailed in \cite{lawrie1989}.  If the scale factor $a(t)$
is also to be determined self-consistently, then there are also two independent
field equations contained in (\ref{fieldequations}). The nature of the solution
is rather difficult to forecast, but we wish to speculate briefly on some
particular features, partly in the light of investigations reported by
Boyanovsky {\it et. al.}  for the somewhat simpler case of the large-$N$
model.  These authors studied a phase transition roughly of the kind envisaged
here in \cite{boyanovsky1997} using a fixed de Sitter background and in
\cite{boyanovsky1998} by solving the Friedmann equation simultaneously with the
field theory problem.  The field-theory equations they solve have, to some
extent, a similar structure to (\ref{sigmaeom}) and (\ref{sigmagap}). There are
trivial differences arising from the fact that, in order to obtain the action
(\ref{action}) (which is equivalent to a Minkowski-space theory with a
time-dependent mass simply proportional to $a(t)$) we have used
$\phi(x)=a(t)\Phi(x)$, where $\Phi(x)$ is a genuine scalar field under general
coordinate transformations, taken this field to be conformally (rather than
minimally) coupled to gravity and used a conformal time coordinate $t$.  There
are, however, more fundamental differences.  Most obviously, the one-loop
equations are exact for the large-$N$ model, whereas those we have given
explicitly are just the lowest non-trivial order of our perturbative
approximation scheme.  In the large-$N$ model, the quantity denoted in
\cite{boyanovsky1998,boyanovsky1997} by $\eta(t)$, which is analogous to our
$\sigma_R(t)$ is the expectation value of one of the $N$ fields.  The evolution
of this field is entirely classical, while the quantum modes contributing to
the self-energy integral $I_1$ arise purely from the $N-1$ transverse fields.
As a consequence, the mode equation corresponding to (\ref{newmode}) is
identical for $k=0$ to the equation of motion for $\eta(t)$, while here it
coincides with (\ref{sigmaeom}) only when $\sigma_{\rm R}(t)$ is small. In the
absence of explicit symmetry breaking, one has $\eta(t)\equiv 0$ and the
calculation is analogous to using our $\phi$-theory for all times. The large-$N$
limit produces, of course, a constrained Gaussian model, in which there is no
scattering, so the dissipative effects treated in section \ref{dissipation} are
entirely absent.

Suppose, then, that our $\phi$-theory were used for all times.  The relevant gap
equation is then (\ref{mugap}) and the mode functions contained in
$\tilde{I}_1^{\phi}(t)$ obey (\ref{newmode}) with $M(t)$ replaced by ${\mu(t)}$.
Initially, $\mu^2$ is positive and $\tilde{I}^{\phi}_1(t)$ should vary only
slowly with time.  Since $a(t)$ increases with time, $\mu^2$ becomes negative,
at which point the mode functions $f_k(t)$ for which $k^2+\mu^2<0$ begin to
grow, and so therefore does $\tilde{I}^{\phi}_1(t)$.  If this growth is fast
enough, then $\mu^2$ might again become positive and, perhaps, a final state
might emerge in which $\mu^2$ remains positive.  In the large-$N$ model, this is
more or less what happens, except that the quantity analogous to $\mu^2$
approaches a vanishingly small value, corresponding to a universe populated
entirely by Goldstone bosons.  In our $N=1$ model, a late-time state with
$\mu^2\ge 0$ would make little sense (though this does not rule it out as a
solution to our 1-loop equations!).  In such a state, the thermal contribution
to the effective mass is still large enough for the symmetry to remain unbroken
(in the conventional sense):  in the (not strictly appropriate) language of
equilibrium field theory, the system would have reheated to above its critical
temperature.  In order to obtain a sensible final state, it is essential to move
to the $\zeta$ description at some point during the growth of the unstable
modes.  From (\ref{sigmaeom}) and (\ref{sigmagap}), it is clear that when
$\sigma_{\rm R}$ is positive, it grows more rapidly than any of the unstable
modes (and the same should be true of $v_{\rm R}$ in our preferred
approximation scheme), so that $M^2$ is more likely than $\mu^2$ to become
positive at late times, as it should.

Finally, we comment briefly on two significant issues raised by the growth of
unstable modes. First, these modes give rise to large propagators which threaten
to make perturbation theory useless.  However, as indicated above, the growth
does not continue indefinitely, and the period of growth will become shorter
if the coupling is made stronger.  In fact, the results presented in
\cite{boyanovsky1997} show a quantity analogous to $\lambda\tilde{I}_1(t)$
growing to a value of 1 in the large-$N$ model and to about 1/3 in a Hartree
approximation which is more similar to our 1-loop approximation.  While these
numbers are not small, they seem to offer hope that perturbation theory may not
become grossly unreliable.  In our $\zeta$-theory, moreover, vertices involving
$\zeta^n$ carry factors proportional to $v^{2-n}$ (see (\ref{lint})).  Since
each factor of $\zeta$ in ${\cal L}_{\rm int}^{\zeta}$ corresponds roughly to
one mode function $f_k(t)$ inside a Feynman diagram, the growth of these mode
functions will be largely offset by the growth of $v_{\rm R}$. The second issue
concerns the nature of the state at late times.  In \cite{boyanovsky1998}, the
small-$k$ modes of the large-$N$ model are interpreted as constituting an
effective classical field
$\phi_{\rm eff}\sim\sqrt{\langle\phi^2\rangle_{\rm R}}$.  Here, this role is
explicitly assumed by $v_{\rm R}$, but the remaining quantum field $\zeta$
still contains the growing modes.  Once these modes have acquired large
amplitudes, the mode equation (\ref{newmode}) provides no mechanism for these
amplitudes to decay, and at first sight, this seems at odds with the
perturbative description of the thermalized state one might expect finally to
emerge. One would expect to describe this state in terms of mode functions with
amplitudes of order $(k^2+M^2)^{-1/4}$, and the presence of large amplitude
modes seems to suggest coherent behaviour that ought to have been absorbed
in the expectation value $v_{\rm R}$.  However, the decomposition of our
propagators $g_{\alpha\beta}$ into mode functions and generalized occupation
numbers $N_k(t)$ is not unique.  At late times, it will be possible to reexpress
these propagators in terms of small-amplitude mode functions and large
occupation numbers for the previously unstable modes.  We then see that the
dissipative mechanisms of section \ref{dissipation} should cause the expected
thermalization (though this does not depend on how we choose to write the
propagators down!).  Indeed, dissipation will be present during the period
of instability as well.  Its effects are difficult to forecast, but some
inhibition of the growth of unstable fluctuations seems likely.


\newpage
\begin{figure}
\caption{Schematic representation of a time dependent effective potential and of
the probability density for $\phi$.}
\label{}
\end{figure}

\begin{figure}
\caption{Diagrammatic representations of the propagators.  Solid lines (a)
represent $g^{\zeta}_{\alpha\beta}$, dashed lines (b) represent $g^{\phi}_{ab}$
and dotted lines are $\phi$ propagators whose time arguments are both equal to
$t_0$.}
\end{figure}

\begin{figure}
\caption{Vertices representing interactions in the $\phi$ and $\zeta$ theories.}
\end{figure}

\begin{figure}
\caption{Low-order diagrams contributing to $\langle\zeta(t)\rangle$.  Anchored
contributions (a) contain composite propagators with vertices fixed at $t_0$
while the free contributions (b) contain none of these vertices.}
\end{figure}

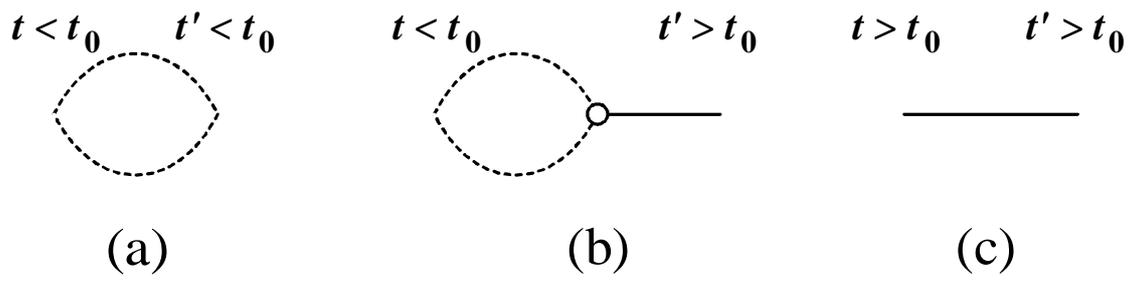
\begin{figure}
\caption{Leading-order contributions to $\langle\phi^2(t)\phi^2(t')\rangle$ when
(a) both times are earlier than $t_0$, (b) one time is later than $t_0$ and
(c) both times are later than $t_0$.}
\end{figure}

\begin{figure}
\caption{Low-order contributions to the connected two-point function
$\langle\zeta_{\alpha}(t)\zeta_{\beta}(t')\rangle$.}
\end{figure}

\end{document}